\documentclass[aps,pre,twocolumn,groupedaddress,showpacs,preprintnumbers,amsmath]{revtex4-1}
\usepackage{epsfig,amsmath,amssymb,graphics,color,calc,hyperref}
\usepackage{braket}
\usepackage{mathtools}

\newcommand{\hg}{\hat{\gamma}}

\begin{document}

\title{A toy model for anomalous transport and Griffiths effects near the Many-Body Localization transition}


\author{M. Schir\'o}\thanks{ On Leave from: Institut de Physique Th\'{e}orique, Universit\'{e} Paris Saclay, CNRS, CEA, F-91191 Gif-sur-Yvette, France}
\affiliation{JEIP, USR 3573 CNRS, Coll\'ege de France,   PSL  Research  University, 11,  place  Marcelin  Berthelot,75231 Paris Cedex 05, France}
\author{M. Tarzia}
\affiliation{LPTMC, CNRS-UMR 7600, Sorbonne Universit\'e, 4 Pl. Jussieu, F-75005 Paris, Franc}

\begin{abstract}
We introduce and study a toy model 
for anomalous transport and Griffiths effects in one dimensional quantum disordered isolated systems near the Many-Body Localization (MBL) 
transitions. The model is constituted by a collection of $1d$ tight-binding chains with on-site random energies, locally coupled to a weak GOE-like perturbation, which mimics the effect of thermal inclusions due to delocalizing interactions by providing a local broadening of the Poisson spectrum.
While in absence of such a coupling the model is localized as expected for the one dimensional Anderson model, increasing the coupling with the GOE perturbation we find a delocalization transition 
to a conducting one driven by the proliferation  of quantum avalanches which does not fit the standard paradigm of Anderson localization. 
In particular an intermediate Griffiths region emerges, where exponentially distributed insulating segments coexist with a few, rare resonances.
Typical correlations decay exponentially fast, while
average correlations decay as stretched exponential and diverge with the length of the chain, 
indicating that the conducting inclusions have a fractal structure and that the localization length is broadly distributed at the critical point.
This behavior is consistent with a Kosterlitz-Thouless-like criticality of the transition.
Transport and relaxation are dominated by rare resonances and rare strong insulating regions, and show anomalous behaviors strikingly similar to those observed in recent simulations and experiments in the bad metal delocalized phase preceding MBL. 
In particular, we find sub-diffusive transport and power-laws decay of the return probability at large times, with exponents that gradually change as one moves across the intermediate region. Concomitantly, the a.c. conductivity vanishes near zero frequency
with an anomalous power-law.
\end{abstract}


\maketitle

\section{Introduction} 

Noninteracting electrons in disordered media display
a uniquely quantum phenomenon known as Anderson
localization~\cite{anderson}; When all electronic states are Anderson
localized, dc transport is absent. Evidence from perturbative~\cite{BAA,Gornyi}, numerical~\cite{Huse}, experimental~\cite{experiments1,experiments2,experiments3}, and rigorous mathematical
approaches~\cite{ImbrieProof} indicate that the main features of Anderson localization (in particular, the absence of diffusion and
dc transport) persist in the presence of interactions. The resulting phase, known as the many-body localized (MBL)
phase~\cite{reviewMBL,reviewMBL2,reviewMBL3,reviewMBL4,reviewMBL5}, has a number of remarkable features: 
a system
in the MBL phase is nonergodic---i.e., its many-body eigenstates violate the eigenstate thermalization hypothesis~\cite{ETH}, the spreading of entanglement is logarithmically slow~\cite{logentanglement}, and common concepts of statistical mechanics break down~\cite{localizationprotected}---and supports extensively many local conserved
quantities~\cite{LIOMS}, which can be constructed using different analytical and numerical approaches~\cite{LIOMS2}.

While there has been a great deal of recent work establishing the existence and properties of the MBL phase (see, e.g., Refs.~\cite{reviewMBL,reviewMBL2,reviewMBL3,reviewMBL4,reviewMBL5} for recent reviews), little is known about the transition between the
MBL and delocalized phases. It is expected that, for sufficiently weak disorder and strong interactions, eigenstates should remain ergodic and transport should be diffusive, as in clean nonintegrable metallic systems~\cite{BAA}. 
However, it has been proposed that diffusivity and/or ergodicity may break down as the MBL transition is approached~\cite{BAA,dot}, even before transport vanishes. Thus, there might be an intermediate phase (called
the ``bad metal'') between the conventional metallic phase and the MBL phase. In this regime the many-body wave-functions might be delocalized but not ergodic and transport is expected to be highly
heterogeneous and strongly fluctuating.

The interest on the delocalized side of the transition started in fact only very recently (see~\cite{BarLev,reviewdeloc1} for recent reviews), when it was observed
that 
in a broad range of parameter before MBL, 
transport is sub-diffusive and out-of-equilibrium relaxation toward thermal equilibrium
is anomalously slow and described by power-laws with exponents that gradually approach zero at the transition.
These features appear as remarkably robust: They were 
observed 
in the numerical solution of the 
the self-consistent BAA equations~\cite{daveBAA}, in numerical simulations (mostly based on exact diagonalizations of samples of moderately small sizes) of disordered spin chains and interacting particles in a random potential~\cite{dave1,BarLev,demler,alet,torres,luitz_barlev,doggen,evers}, as well as 
in recent experiments with cold atoms~\cite{experiments1,experiments2,experiments3}.

An appealing phenomenological interpretation of these phenomena has been proposed in terms of the existence of a quantum Griffiths phase~\cite{GP,Vojta}.
The idea is that a system close to MBL is highly inhomogeneous (in real space) and is characterized by rare inclusions of the
insulating phase with an anomalously large escape time (i.e., anomalously small localization length). 
In $1d$ such insulating segments affect dramatically the dynamics, since quantum excitations have to go through broadly distributed 
effective barriers which act as kinetic bottlenecks and give rise to sub-diffusion, slow relaxation, and anomalous spectral correlations~\cite{vosk,potter1,potter2,demler,dave1,griffiths2,reviewdeloc1,serbyn_moore}, in a way which is very similar to the trap model for glassy dynamics~\cite{trap}.

The first phenomenological descriptions of the sub-diffusive ergodic phase in terms of Griffiths regions was proposed in Ref.~\cite{demler}, in terms of a 
classical resistor-capacitor model with power-law distributed resistances~\cite{RC}, and similar classical trap-like models~\cite{BarLev,griffiths2}.
Later, Griffiths effects have been investigated within strong-randomness Renormalization Group (RG) approximations ~\cite{vosk,potter1,zhang,potter2,thiery1,thiery,goremykina,dimitrescu,morningstar},
devised to investigate the asymptotic critical behavior
of the MBL transition. 
These works 
provided numerically implemented RGs designed to
capture the physics of interactions between locally thermal and MBL regions.

 Despite being based on the same idea of coarse-graining many-body resonances in a strong disorder approach, these various proposed RGs differ in the way in which the thermal and insulating regions are identified and combined during the RG steps. The assumptions behind these constructions, which are all essentially phenomenological, are motivated in part by the requirement that the MBL transition itself must be universal (i.e., independent on the microscopic details) and thermal (i.e., incoherent and classical). Accordingly, one could expect that the critical properties should be well described by an effective classical statistical mechanics model. 

On a different front, the idea that the MBL/ETH transition could be driven by quantum avalanches has been recently put forward in~\cite{thiery,thiery1,avalanches} by studying the way
a localized system react to coupling with a thermal bath~\cite{altman}. According to this picture, the MBL phase may be destabilized by finite  ergodic ``bubbles'' of weak disorder 
that occur naturally inside an insulator and that may trigger a ``thermalization avalanche''. This mechanism has also found support in the latest RG
studies~\cite{goremykina,dimitrescu,morningstar}, which have shown that the avalanche process combined with a natural choice of the scaling variables immediately leads to  
a Kosterlitz-Thouless (KT) critical behavior for the MBL transition.




In this light, it would be desirable to have a tractable quantum model for Griffiths effects which reproduces the critical behavior predicted by the phenomenological RG approaches
and yet retains full quantum mechanical nature, thereby allowing to study, for example, the Griffiths signatures in the quantum dynamics.
To this aim 
in this paper we develop and study a new 
microscopic quantum toy model for Griffiths effects, anomalous transport and relaxation 
close to the MBL transition.
The model is built on random matrices and is analytically tractable.
From one side, the model is inspired by the minimal effective coarse-grained descriptions of Refs.~\cite{vosk,potter1,potter2,thiery,thiery1} designed to capture the essence of the MBL transition and the formation of many-body resonances in the framework of the strong disorder RG approach. 
On the other hand, the model is designed to study how a Anderson insulator can react to coupling with a thermalizing system~\cite{altman} (see also Ref.~\cite{zeno}), leading to the picture of quantum avalanches~\cite{avalanches}.

We show that the model reproduces most of the key features of the bad metal Griffiths phase~\cite{BarLev,reviewdeloc1}, including exponentially distributed localized segment, delocalization due to quantum avalanches produced by a fractal set of thermal inclusions, broadly distributed localization lengths, 
sub-diffusion, and anomalous power-law transport and relaxation.

The paper is organized as follows. In the next section we introduce the model; 
In Sec.~\ref{sec:recursion} we derive the exact recursion relations for the Green's functions and give the intuitive arguments for the formation of resonances responsible for the delocalization transition; In Sec.~\ref{sec:ploc} we analyze the metal/insulator transition and draw the phase diagram; In Sec.~\ref{sec:griffiths} we investigate the properties of the critical region; 
In Sec.~\ref{sec:dynamics} we focus on the sub-diffusive dynamics and anomalously slow power-law transport and relaxation observed in the intermediate regime;
In Sec.~\ref{sec:1deff} we rewrite the toy model as an effective one dimensional problem. 
The result can be seen either as an effective Anderson model with correlated and self-consistently generated disorder or as an effective Anderson Hamiltonian in presence of many-body interactions, resulting in a non trivial self-energy correction to the local Green's function.
Finally, in Sec.~\ref{sec:conclusions} we give some concluding remarks and perspectives for future works. Further details and information are given in Apps.~\ref{app:eliminating}-\ref{app:resist}.    

\section{The model} \label{sec:model}

Several effective descriptions for the formation of collective many-body resonances that destabilize the MBL phase 
have been proposed in the literature in the latest years in the context of the strong disorder RG approach to MBL~\cite{vosk,potter1,zhang,potter2,thiery,thiery1}.
These models are generic coarse-grained one-dimensional models with short-ranged interactions and no specific microscopic structure, built on random matrices.
The basic assumption behind these constructions is that 
sufficiently close to the critical point one can consider an
effective model in terms of resonant clusters, i.e., groups of inter-resonating single-particle orbitals, characterized only
by coarse grained information and a minimal set of parameters~\cite{vosk,potter1,potter2,thiery,thiery1}. 

\begin{figure}
\includegraphics[width=0.48\textwidth]{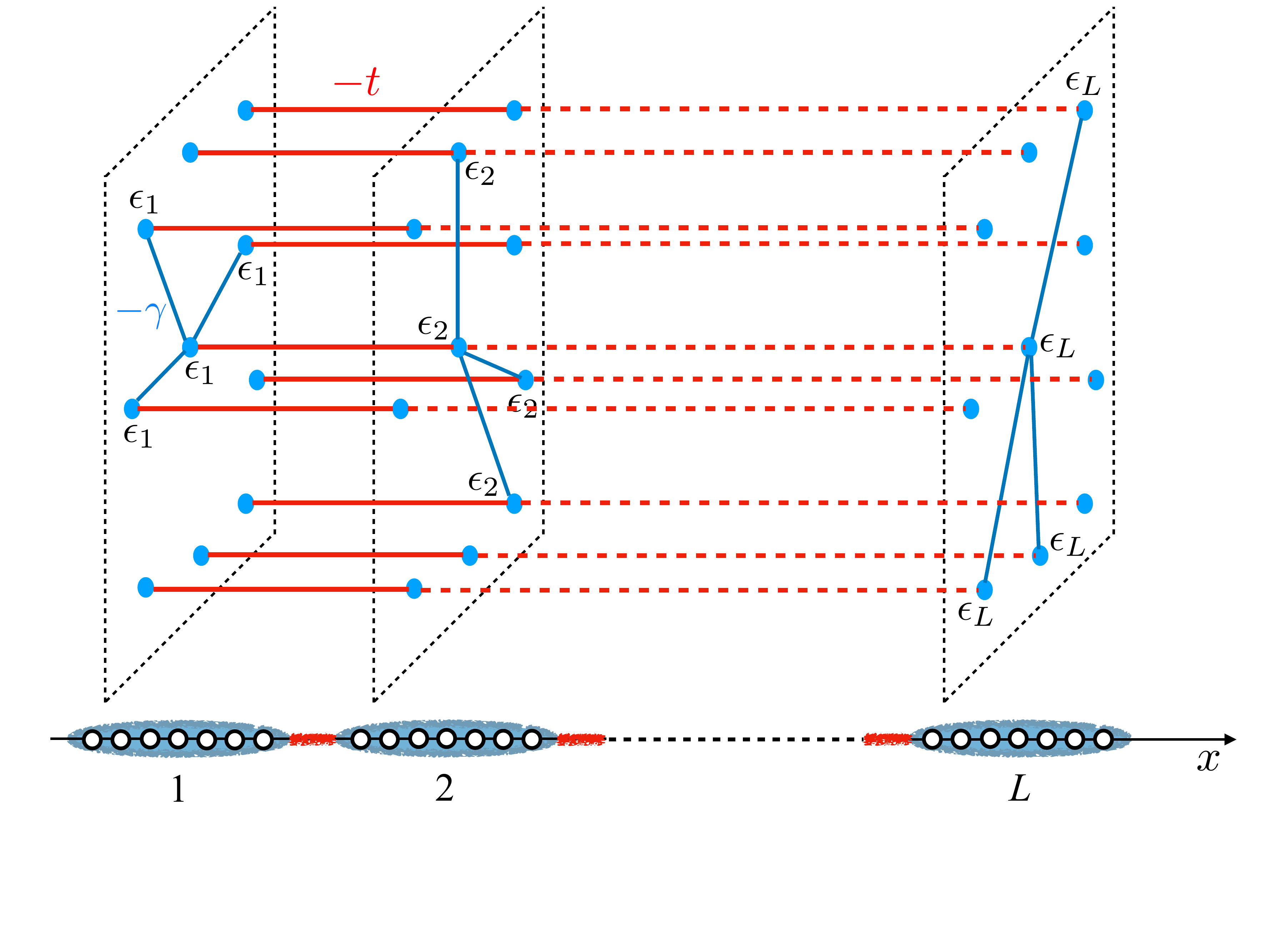}%
\vspace{-0.9cm}

\includegraphics[width=0.48\textwidth]{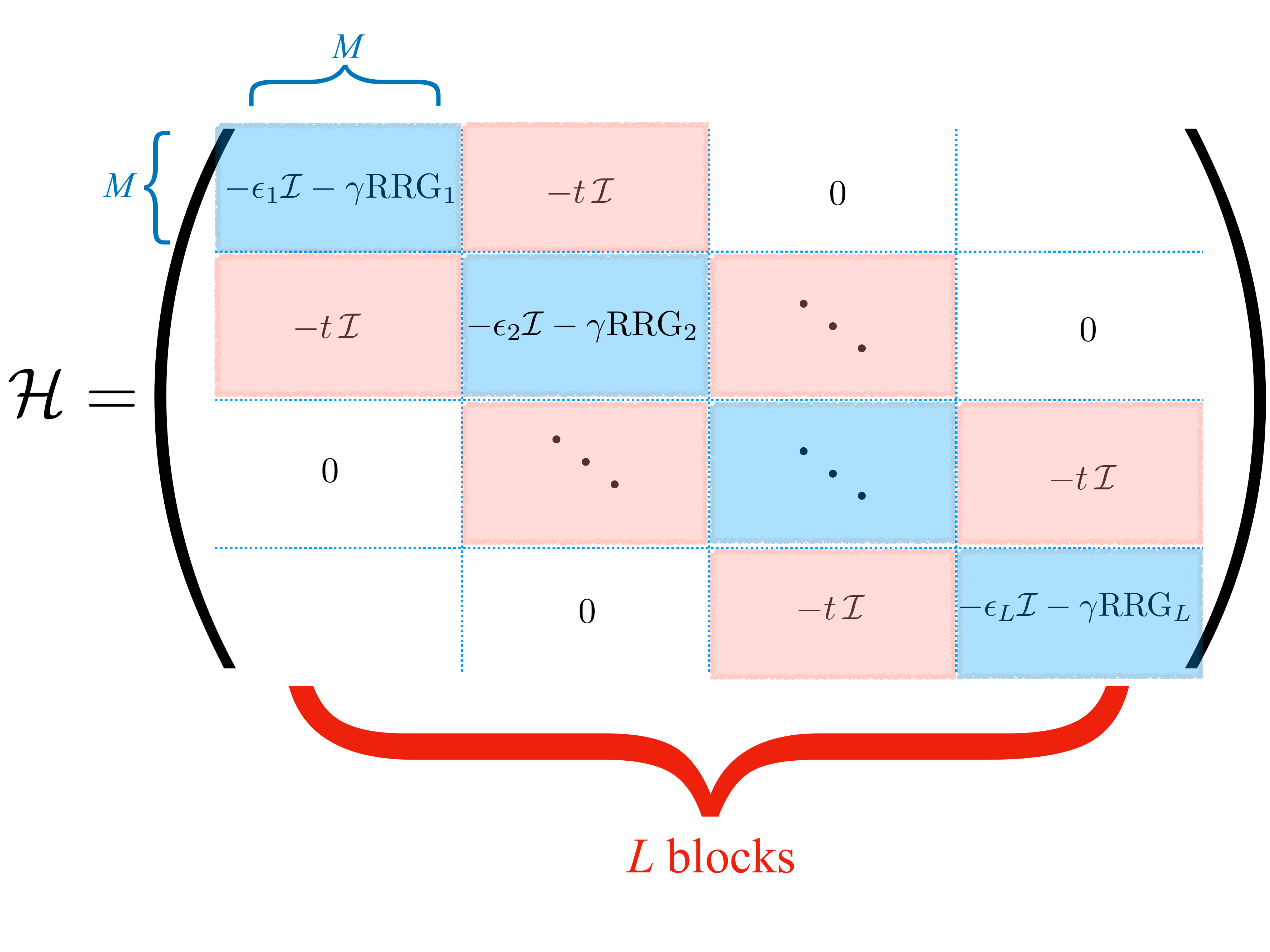}%
\caption{\label{fig:model}
        Pictorial representation of the model~(\ref{eq:H}). Top: The model is made of a $1d$ chain of $L$ layers coupled along the $x$-direction via the inter-layer hopping rates $t$, with random energies $\epsilon_i$ (equal on all sites of 
        the layer) extracted from a box distribution of width $W$. Each layer contains $M$ sites. The matrix elements between two sites belonging to a given layer correspond to the adjacency matrix of a random realization of a RRG 
        (different on each layer) times the intra-layer hopping rate $\gamma$. 
        One can imagine that each layer of the chain is a pictorial representation 
        of a coarse-grained block of $\ell$ sites of an interacting many-body problem, 
        and that 
        the $M \times M$ RRG matrices are an extreme simplified description of the Hilbert space of the local degrees of freedom on each segment in some specific basis.
        Bottom: The matrix representation of the Hamiltonian consists in $L$ blocks of $M \times M$ matrices. The diagonal blocks are ($- \gamma$ times) random realizations of the connectivity matrix of a RRG, shifted by random energies $\epsilon_i$. Adjacent blocks are connected by blocks of the form $- t {\cal I}$.}
        \vspace{-0.3cm}
\end{figure}

Here we introduce a toy model inspired by these approaches: 
We consider $M$ identical copies, labeled by the index $p=1, \ldots, M$, of a $1d$ Anderson tight-binding model on $M$ chains of length $L$; At each horizontal position $i$, the $M$ sites belonging to different 
chains are coupled by random hopping terms of strength $\gamma$ extracted from a 
sparse random matrix ensemble, 
i.e., the ensemble of random-regular-graphs (RRG) of fixed total connectivity $k+1$~\cite{RRG} (we will set $k+1 = 3$ hereafter). 
RRG are random lattices which have locally a tree-like structure but have loops whose typical length scales as $\log M$ and no boundary, and are statistically translationally invariant.
The Hamiltonian of the model is:
\begin{equation} \label{eq:H}
\begin{aligned}
{\cal H} =& - \sum_{i=1}^L \Bigg \{  \sum_{p=1}^M \left[ \epsilon_i \, d^\dagger_{i,p} d_{i,p} 
+ t 
 \left ( d^\dagger_{i,p} d_{i+1,p} + \textrm{h.c.} \right) \right] \\ 
& \qquad \qquad - \gamma 
\sum_{\langle p,q \rangle_i} \left ( d^\dagger_{i,p} d_{i,q} + \textrm{h.c.} \right) \Bigg \} \, , 
\end{aligned}
\end{equation}
where $d_{i,p}$ and $d^\dagger_{i,p}$ are creation and annihilation operators on the site $p$ of the $i$-th layer, 
$\epsilon_i$ are i.i.d. random energies taken uniformly from a box distribution on $[-W/2,W/2]$ (which, for simplicity, we take identical on all sites $p$ sitting at the same position $i$ of the chains), 
$t$ is the inter-layer hopping rate between sites belonging to adjacent layers, and $\gamma$ is the intra-layer hopping rate between sites $p$ and $q$ with the same horizontal coordinate $i$. The notation $\langle p,q \rangle_i$
indicates couples of sites connected by a link within the $i$-th layer. Note that on each layer a different random realization
of the RRG is chosen, in such a way that two sites that are connected by $\gamma$ within a given layer are (with high probability in the $M \to \infty$ limit) not connected on the other layers. This is important as it ensures that the whole lattice 
can be thought as an {\it anisotropic} random graph of total connectivity $k+3$, which is locally a tree but has loops whose typical size scales as the system size. A pictorial representation of such lattice is given in Fig.~\ref{fig:model}.

It is known from previous studies that the RRG ensemble of sparse random matrices belongs to the GOE universality class (with Wigner-Dyson-level statistics and fully delocalized eigenvectors)~\cite{RRG-GOE,Bauerschmidt}. 
Hence, in absence of the hopping rates connecting sites on adjacent layers ($t=0$), each layer $i$ corresponds to a $M \times M$ GOE-like block, with energy spectra akin to semicircle laws of width $4 \gamma \sqrt{k}$~\cite{ourselves} and centered around $\epsilon_i$. When the inter-layer hopping 
matrix elements is turned on ($t>0$), these GOE-like blocks become then coupled along the chain. The local GOE-like perturbation thus mimics the effect of the delocalizing interaction,
thereby allowing to study, within the framework of a tractable quantum toy model, the competition between localization and thermalization.

To connect with other phenomenological models for many-body resonances, 
one might imagine that each layer $i$ of the chain 
represents a coarse-grained block of $\ell$ sites of an interacting many-body system 
and that 
the $M \times M$ RRG matrices 
are as an extreme simplified description of the Hilbert space of the local degrees of freedom on each segment
(with $M \sim e^\ell$) in some specific basis (e.g., the Fock space).
Within this interpretation the effective degrees of freedom in Eq.~(\ref{eq:H}) 
should be in
fact thought as local many-body quasiparticle excitations of the interacting systems within the coarse-grained blocks. 
The hopping rate $t$ thus plays the role of the entanglement rate between 
energy levels of adjacent blocks. 
In a truly interacting many-body problem the Hilbert space is the tensor product of the Hilbert space of local degrees of freedom and its dimension should scale as $M^L$, 
differently from the non-interacting toy model introduced here, for which the total size of the Hilbert is only $M L$. A possible justification of that is given by the observation that our toy model 
should be thought as a pictorial description of the transition from the MBL phase to the thermal one coming from the former. MBL eigenstates are exponentially
localized in the Hilbert space and exhibit short-range entanglement that scales as the perimeter of the coarse-grained blocks. It is thus reasnoable to assume that keeping only a small portion
of the total Hilbert space that grows linearly with the number of blocks might provide a plausible starting point for a zero-th order simplified description.
In this sense, our model is very similar in spirit to the effective coarse-grained models introduced and studied in the context of the strong disorder RG approach to MBL~\cite{vosk,potter1,potter2,thiery,thiery1} (see also Ref.~\cite{zeno}). 
Yet, it is a non-interacting tight-binding model for spinless electrons on a tree-like (although anisotropic) random lattice, and it can be solved exactly and its properties can be studied analytically in full details.
The model~(\ref{eq:H}) is also 
a $M$-orbital version of the $1d$ Anderson model, which can be mapped onto a supersymmetric $\sigma$-model~\cite{nlsm} (however, in general one assumes Gaussian distributed random matrix elements).

\section{Exact recursion relations} \label{sec:recursion}

\begin{figure}
\includegraphics[width=0.48\textwidth]{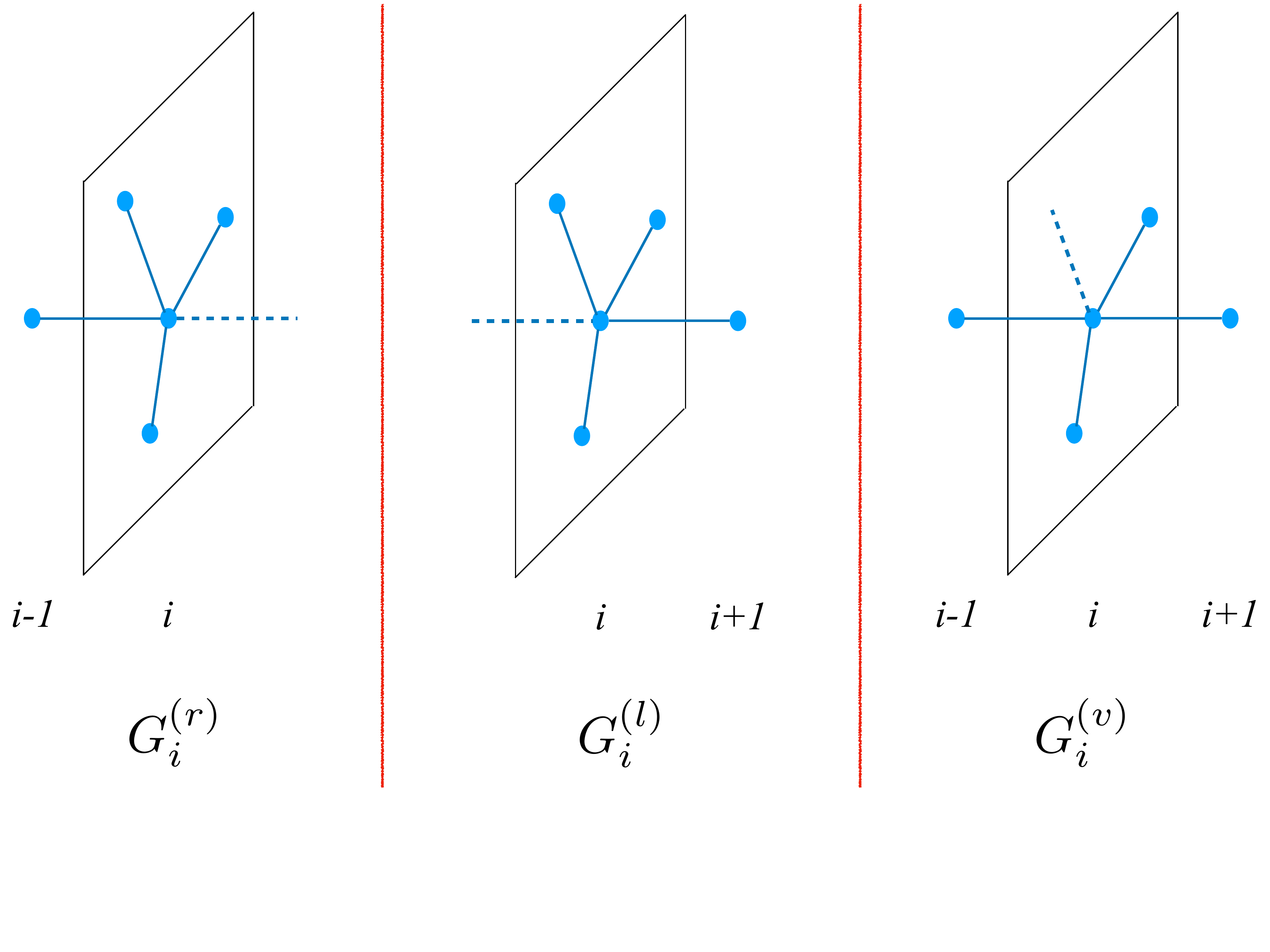}
\caption{\label{fig:cavity}
Sketch of the three kinds of cavity Green's functions defined on the anisotropic RRG.}
\vspace{-0.5cm}
\end{figure}

The model~(\ref{eq:H}) allows, in principle, for an exact solution which yield the probability distribution function of the diagonal elements of the
resolvent matrix, defined as ${\cal G} (z) =  ({\cal H} - z {\cal I}  )^{-1}$~\cite{abou,ourselves}.
In order to obtain the recursive equations, the key objects are the so-called {\it cavity} Green's functions,
i.e., the diagonal elements on a given site $i$ of the resolvent matrix of the modified Hamiltonian where the edge between
the site $(i,p)$ and one of its neighbors has been removed. 
Due to the anisotropic structure of the lattice, we need to define three kinds of cavity Green's functions, 
respectively in absence of a link between site $(i,p)$ and its left neighbor $(i-1,p)$, $G_{i,p}^{(l)}$, its right neighbor $(i+1,p)$, $G_{i,p}^{(r)}$, and one of the $k+1$ neighbors $(i,q)$ belonging to the same layer, $G_{i,p}^{(v)}$ (see Fig.~\ref{fig:cavity} for a sketch).
For simplicity in the following we will take the limit $M \to \infty$ from the start, although the results are essentially unchanged for large but finite values of $M$ (provided that $M \gg L$).
The advantage of taking $M \to \infty$ is twofold: First, in absence of intra-layer disorder, all sites belonging to a given layer become 
equivalent and the cavity Green's functions 
become translationally invariant within each layer (i.e., they are identical on all sites $p$ of the $i$-th layer, $G_{i,p}^{(l,r,v)} = G_{i}^{(l,r,v)}~\forall p$);
Second, since the system is already of infinite size (at least formally), one can take te limit $\eta = 0^+$ from the start.
Thanks to the tree-like structure of the graph, in the $M \to \infty$ limit the following iteration relations 
can be easily obtained (e.g., by Gaussian integration):
\begin{equation} \label{eq:recursion}
\begin{aligned}
\left[G_{i}^{(l)}\right]^{-1} &= - \epsilon_i - z - t^2 G_{i+1}^{(l)} - (k+1) \gamma^2 G_{i}^{(v)}  \, , \\
\left[G_{i}^{(r)}\right]^{-1} &= - \epsilon_i - z - t^2 G_{i-1}^{(r)} - (k+1) \gamma^2 G_{i}^{(v)}  \, , \\
\left[G_{i}^{(v)}\right]^{-1} &= - \epsilon_i - z - t^2 G_{i+1}^{(l)} - t^2 G_{i-1}^{(r)} - k \gamma^2 G_{i}^{(v)}  \, , \\
\end{aligned}
\end{equation}
where $z=E + i \eta$, $\eta \to 0^+$ is an infinitesimal imaginary regulator, 
$\epsilon_i$ are the on-site random energies.
Once the solution of Eqs.~(\ref{eq:recursion}), which is a system of $3 L$ coupled non-linear equations, has been found, one can finally obtain the diagonal
elements of the resolvent matrix of the original problem on a given site 
as a function of the cavity Green's functions on the neighboring sites~\cite{ourselves}:
\begin{equation} \label{eq:recursion_final}
{\cal G}_i = \frac{1}{- \epsilon_i - z - t^2 G_{i+1}^{(l)} - t^2 G_{i-1}^{(r)} - (k+1) \gamma^2 G_{i}^{(v)} } \, .
\end{equation}
The last term in the denominator of the previous expression, $- (k+1) \gamma^2 G_{i}^{(v)}$, represents the correction to the Green's functions due to the GOE-like perturbation with respect to the bare $1d$ Anderson model, and might be interpreted as the correction that one would obtain by treating the interacting term of a $1d$ many-body Hamiltonian using some kind of self-consistent approximation (as we show more explicitly in Appendix ~\ref{app:eliminating}, see, e.g., Eqs.~(\ref{eq:recursion_sigma}) and~(\ref{eq:sigma_corrections})). Since the term $- (k+1) \gamma^2 G_{i}^{(v)}$ is frequency-dependent, the structure of the equations corresponds to a correction which goes beyond the Hartree-Fock level~\cite{Weidinger18}, which is purely local in time, and is instead reminiscent of a DMFT-like approximation~\cite{DMFT} within the nonequilibrium Keldysh field theory formalism~\cite{noneqDMFT}. Quite interestingly the effect of the local GOE perturbation on the one dimensional problem is also reminiscent of SYK model~\cite{SYK}  and its finite dimensional extensions~\cite{SYK2}, recently proposed to study transport in bad metal phases.

The statistics of the diagonal elements of the resolvent gives---in the $\eta \to 0^+$ limit---the spectral properties of $\mathcal{H}$.
In particular, the probability distribution of the Local Density of States (LDoS) at energy $E$ is given by:
\begin{equation} \label{eq:LDoS}
        \begin{aligned}
                \rho_i (E) 
                & = \sum_\alpha | \langle i \vert \alpha \rangle |^2 \, \delta ( E - E_\alpha ) 
                =\lim_{\eta \to 0^+}  \frac{{\rm Im} {\cal G}_i (z)}{\pi} \, ,
        \end{aligned}
\end{equation}
from which the average Density of States (DoS) is simply obtained as $\rho (E)
 = (1/L) \sum_i \rho_i (E)$.
  
In the following we will (mostly) focus on the middle of the spectrum ($E=0$) and set $t=1$. From now on we will also consider periodic boundary conditions, 
but the results are unchanged for open chains provided that $L$ is sufficiently large.

\subsection{Intuitive arguments for the formation of resonances} \label{sec:intuitive}

\begin{figure}
\includegraphics[width=0.48\textwidth]{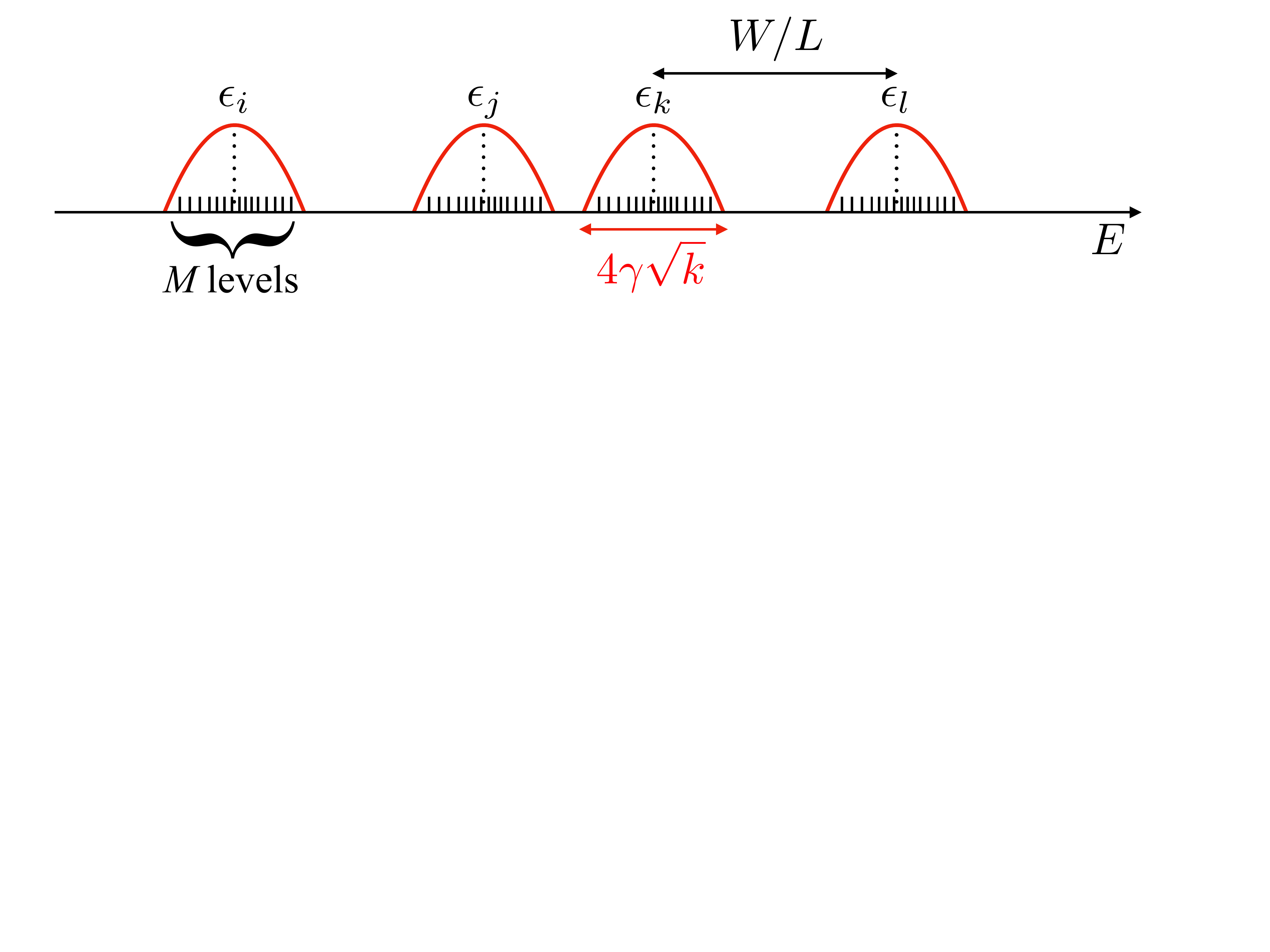}%
\vspace{-4.3cm}
\caption{\label{fig:resonances}
        Pictorial representation of the energy spectrum of the model in the strong disorder limit, $W/t \gg 1$. The DoS is given by a superposition of $L$ semicircles (containing $M$ levels each) centred roughly around the random chemical 
        potentials $\epsilon_i$ and of width $4 \gamma \sqrt{k}$.}
\end{figure}

Increasing the coupling with the GOE-like perturbation $\gamma$ the model undergoes a localization/delocalization transition from an insulating to a conducting phase.
The physical mechanism behind the transition can be understood intuitively from the pictorial sketch of Fig.~\ref{fig:resonances} and goes as follows. 
For sake of simplicity we start by discussing the strong disorder limit, $W/t \gg 1$. In absence of the GOE coupling ($\gamma=0$) the system is composed of $M$ 
identical copies of $1d$ Anderson localized chains. At strong disorder 
eigenfunctions are exponentially localized around specific sites of the chain, the ($M$-degenerate) eigenenergies are only weak modification 
of the on-site random chemical potentials $\epsilon_i$, 
and are Poisson-distributed, the typical distance between two consecutive energy levels being $W/L$.
As soon as the intra-layer perturbation is turned on ($\gamma>0$), the RRG couplings lift the degeneracy by providing an effective broadening of the $M$-degenerate Poisson unperturbed levels. The energy spectrum is thus composed by a 
superposition of $N$ small semicircles (containing $M$ energy levels each) with support (roughly) in the interval $[\epsilon_i-2 \gamma \sqrt{k},\epsilon_i+2 \gamma \sqrt{k}]$. 
One then naturally expects that if $4 \gamma \sqrt{k} \gg W/L$, 
the support of the semicircles superpose, resonances are typically formed between unperturbed states, and ``particles'' delocalize over the whole chain. 
Conversely, if $4 \gamma \sqrt{k} \ll W/L$ the probability of finding a resonance decays exponentially with the distance and 
the ``particles'' stay localized.
The transition is thus expected to occur for 
\begin{equation} \label{eq:edges}
\gamma_c \sim  \frac{W}{4 \sqrt{k} L} \, .
\end{equation}

Another way to understand the mechanism at the origin of the localization/delocalization transition is provided by the locator expansion~\cite{anderson}.
In fact, in absence of the intra-layer coupling ($\gamma=0$) the Green's function element between
a point $i$ and a point $j$ of the chain can be formally expressed as:
\[
{\cal G}_{ij} = \sum_{\cal P} \prod_{l \in {\cal P}} \frac{t}{\epsilon_l} \, , 
\]
where the sum is over all paths ${\cal P}$ connecting $i$ and $j$
and the product is over all sites $l$ belonging to the path.
In $1d$ the weight of a path will decrease exponentially with its length. The sum over
paths will then be dominated by the forward-scattering paths:
\[
{\cal G}_{ij} \approx \prod_{l=i}^j \frac{t}{\epsilon_l} \, .
\]
As soon as the GOE coupling is turned on, a huge number of new directed paths between sites $i$ and $j$ are generated, since at each position $l$ of the chain a ``particle'' can travel $n_l$ steps within the $l$-th RRG before jumping to the 
adjacent one. Since the number of paths of length $n$ on a tree scales as $k^n$, one has that:
\begin{equation} \label{eq:LE}
{\cal G}_{ij} \approx \sum_{ \{ n_l \} } \prod_{l=i}^j \left( \frac{k \gamma}{\epsilon_l} \right)^{n_l} \frac{t}{\epsilon_l} \, .
\end{equation}
Comparing this expression to the $1d$ unperturbed case ($\gamma=0$), the effect of the coupling $\gamma$ is to ``renormalize'' the bare random energies $\epsilon_l$ as
\[
\epsilon_l \to \left( \frac{\epsilon_l}{k \gamma} \right)^{n_l} \, .
\]
For the sites such that $\epsilon_l$ is sufficiently close to $E=0$ (i.e., $|\epsilon_l| < k \gamma$) this can create arbitrarily large terms in the sum~(\ref{eq:LE}). Hence, the effect of the intra-layer coupling $\gamma$
is to enhance resonances, thereby possibly making the locator expansion diverge. This argument suggests that delocalization
is likely to be driven by few rare resonances that may form for some specific realizations of the disorder, and is consistent with the avalanche mechanism as a possible scenario for the MBL transition~\cite{avalanches,thiery,thiery1,dimitrescu,morningstar}: A fractal set of measure zero of thermal inclusions can be enough to thermalize the whole system.
For a chain of length $L$ one expects that the transition occurs when the probability of finding at least one layer with 
$|\epsilon_l| < k \gamma$ becomes of order one:
\begin{equation} \label{eq:FSA}
\gamma_c \sim \frac{W}{2 k L} \, .
\end{equation}
Both arguments indicate that the transition takes place for $\gamma$ of order $1/L$, Eqs.~(\ref{eq:edges}) and~(\ref{eq:FSA}).
For this reason we introduce a new control parameter $\phi$ which tells us how fast the intra-layer coupling decreases with the length of the chain: 
\begin{equation} \label{eq:gammahat}
\gamma = \hat{\gamma}/L^\phi \, ,
\end{equation}
with $\hg$ of order one, 
reminiscent of the unconventional scaling recently proposed to access the many body localization transition in dimensions greater than one or with long-range interactions~\cite{sarang}. 
The transition is expected to occur at $\phi=1$, at least in the strong disorder limit. 

\section{The metal/insulator transition and the phase diagram} \label{sec:ploc}

\begin{figure}
\includegraphics[width=0.48\textwidth]{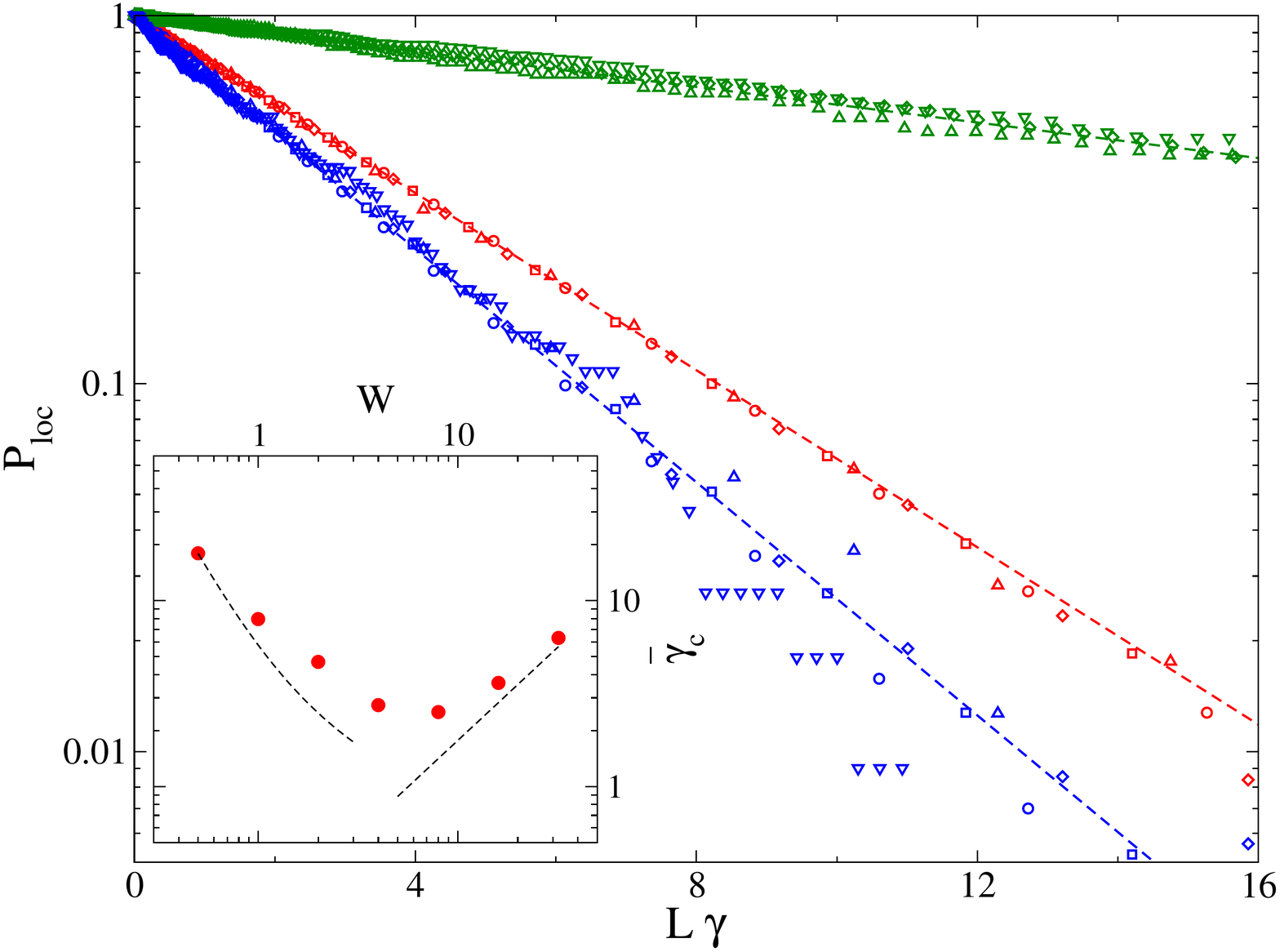}%
\caption{\label{fig:ploc}
Main panel: Probability that a chain of length $L$ is localized, $P_{\rm loc}$ (i.e., the probability that the perturbative series in $\gamma^2$ converges), 
as a function of the GOE coupling $\gamma$, multiplied by the system size $L$, for $L=2^{10}$ (circles), $L=2^{12}$ (squares), $L=2^{14}$ (diamonds),
$L=2^{16}$ (up triangles), $L=2^{18}$ (down triangles), and for three values of the disorder strength, $W=0.5$ (green), $W=4$ (blue), and $W=16$ (red). Inset: Non-monotonic behavior of $\hg_c (W)$ as a function of $W$. The dashed black curves show the asymptotic behavior 
$\hg_c(W) \approx W/( 4 /\sqrt{k})$ at large disorder and  $\hg_c(W) \approx 1/ (W^2 \sqrt{k})$ at small disorder.}
\end{figure}

For $\gamma=0$ Eqs.~(\ref{eq:recursion}) and~(\ref{eq:recursion_final}) reduce to the recursion relations for the Green's functions of the $1d$ tight-binding Anderson model, i.e., they are unstable 
with respect to the imaginary regulator $\eta$ for any positive value of $W$: $P({\cal G})$ is singular and the average Density of States (DoS) vanishes in the $\eta \to 0^+$ limit.
When the GOE perturbation is turned on, the Green's functions can be systematically expanded in powers of $\gamma^2$: The delocalization transition occurs at the value of $\gamma$ at which such perturbative expansion is not convergent, 
implying that a stable non-singular probability distribution of ${\rm Im} {\cal G}$ is generated by the intra-layer coupling.
In fact, 
as shown more in detail in Sec.~\ref{sec:1deff} and App.~\ref{app:self}, the GOE perturbation plays essentially the role of a thermal bath, by introducing on each layer $i$ a local small (i.e., of order $\gamma$) source of dissipation. In order to understand whether such dissipation propagates throughout
the whole system, 
one can solve the recursion relations 
order by order, 
and check whether the perturbative series in $\gamma^2$ converges.
The results are shown in Fig.~\ref{fig:ploc}, where we plot the probability $P_{\rm loc}$ that a system of length $L$ stays localized for a given disorder realization, 
for several system sizes and three values of $W$.
We find that $P_{\rm loc}$ decays exponentially with $\gamma$ and the curves for different $L$ nicely collapse on the same function when $\gamma$ is multiplied by the system size:
\begin{equation} \label{eq:Ploc}
P_{\rm loc} = e^{- L \gamma / \hg_c (W)}  = e^{- \hg / \hg_c (W)} \, .
\end{equation}
The presence of exponentially rare large localized regions is precisely the 
hallmark of the Griffiths phase expected to describe the delocalized side of MBL systems close enough to the transition~\cite{reviewdeloc1}. These
inclusions act as bottlenecks~\cite{dave1,BarLev,demler,griffiths2,reviewdeloc1}, leading to sub-diffusive transport and sub-ballistic spreading of entanglement.

Quite interestingly, 
the disorder-dependent characteristic scale of the intra-layer coupling on which delocalization takes place, $\hg_c (W)$, 
has a strong non-monotonic dependence on the disorder strength $W$~\cite{zeno}, as shown in the inset of Fig.~\ref{fig:ploc}. The behavior of $\hg_c (W)$ at strong disorder can be understood 
from the argument given in Sec.~\ref{sec:intuitive}, leading to Eq.~(\ref{eq:edges}), i.e.
$\hg_c \approx W/(4 \sqrt{k})$ (dashed curve of the inset of Fig.~\ref{fig:ploc} at large $W$).
Conversely, at weak disorder, the localization length $\xi_0$ of the unperturbed $1d$ chains is very large. The system can be thought as effectively composed by $L/\xi_0$ insulating blocks of average size $\xi_0$, with finite average DoS is in the interval $[-2t,2t]$.
Once the intra-layer perturbation is turned on, one expects that different blocks can amix provided that the effective width provided by the coupling $\gamma$ is of the order of the inverse of the typical distance between the insulating blocks, 
which gives $\hg_c \approx \xi_0 t/\sqrt{k}$ (dashed curve of the inset of Fig.~\ref{fig:ploc} at small $W$).

\begin{figure}
\includegraphics[width=0.48\textwidth]{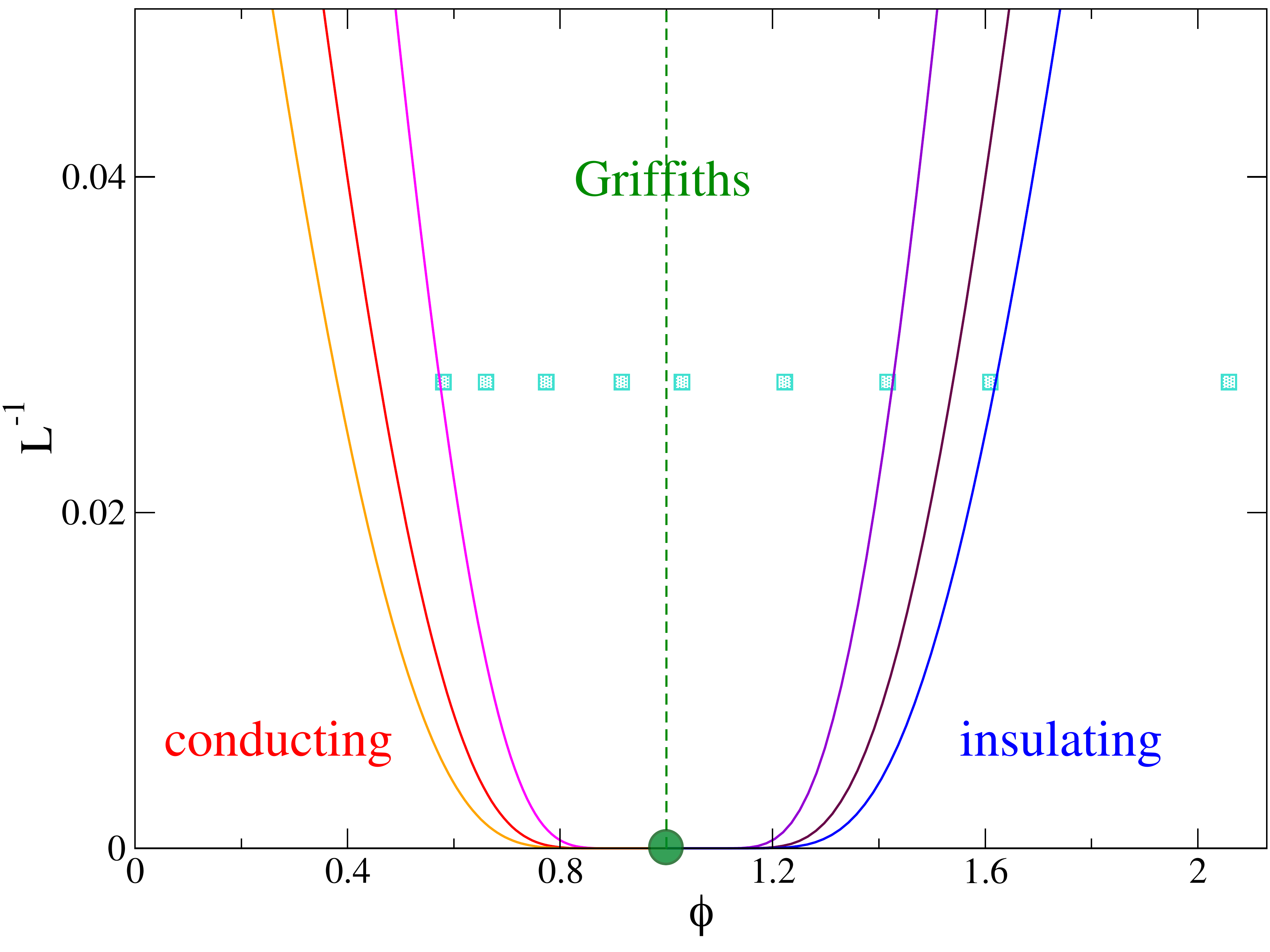}%
\caption{\label{fig:phase_diagram}
Phase diagram of the model in the plane $\phi$-$L^{-1}$. For infinite chains with probability $1$ the system is insulating for $\phi>1$ and conducting for $\phi<1$. At finite $L$ a broad intermediate region appears where broad insulating and conducting segments
coexists. The continuous lines correspond to $P_{\rm loc} \le \epsilon$ and $P_{\rm loc} \ge 1-\epsilon$ respectively, with $\epsilon=10^{-2}$, $10^{-3}$, and $10^{-4}$. The 
light blue squares represents the points where we performed exact diagonalizations (see Sec.~\ref{sec:dynamics}).}
\end{figure}

Introducing the rescaled variable $\hat{\gamma}$ defined in Eq.~(\ref{eq:gammahat}), we can rewrite Eq.~(\ref{eq:Ploc}) as:
\[
P_{\rm loc} (L, \phi) = e^{- (\hg / \hg_c) L^{1-\phi}} \, ,
\]
from which we can draw the
phase diagram 
in the plane $\phi$-$L^{-1}$, shown in Fig.~\ref{fig:phase_diagram}. In the $L \to \infty$ limit $P_{\rm loc}=1$ for $\phi>1$ (the system is localized) and $P_{\rm loc}=0$ for $\phi<1$ (the system is
an conducting). Yet, at finite $L$ there exist a broad intermediate region where arbitrarily large (i.e., of order $L$) metallic and conducting segments coexist, 
and chains of $L$ layers can be either insulating or conducting with a probability between $0$ and $1$
depending on the particular disorder realization.
In Fig.~\ref{fig:phase_diagram} we plot the lines where $P_{\rm loc} (L,\phi) \le \epsilon$ on the delocalized side of the phase diagram, and the lines where
$P_{\rm loc} (L,\phi) \ge 1-\epsilon$ on the localized side (with $\epsilon=10^{-2}$, $10^{-3}$, and $10^{-4}$), showing that the intermediate region becomes broader and broader as $L$ is decreased.
For $L \to \infty$ the intermediate region shrinks to a point, $\phi=1$. Yet $P_{\rm loc}$ can be still continuously varied from zero to one by tuning $\hat{\gamma}$ from zero to infinity.

The phase diagram of the model shares some similarities with that of the power-law random banded matrix (PLRBM) model~\cite{PLRBM}, which 
describes particles in $1d$ with random long range hopping. 
The hopping amplitudes between two sites $i$ and $j$ of the chain are i.i.d. Gaussian random variables with zero mean and variance decaying with the distance $|i-j|$ as
$\overline{({\cal H}^{\rm PLRBM}_{ij})^2} = [1 + (|i - j|/b)^{2 \alpha}]^{-1}$. The PLRBM model undergoes an Anderson transition at $\alpha=1$ from the localized
to the delocalized phase for an arbitrary value of $b$ and shows all key features of the finite dimensional Anderson critical point, including multifractality of eigenfunctions and nontrivial spectral
compressibility. In fact the parameter $b$ defines a whole family of critical theories: $b \gg 1$ represents a regime of weak multifractality, analogous to the conventional Anderson transition in
$d= 2 + \epsilon$, while $b \ll 1$ is characterized by strongly fluctuating eigenfunctions, similar to the Anderson transition in $d \gg 1$ (and is accessible to an analytical treatment using a strong-disorder real-space RG 
method~\cite{levitov}). In a certain sense, our exponent $\phi$ plays the role of the exponent $\alpha$ of the PLRBM, while the parameter $\hat{\gamma}$, which allows to tune $P_{\rm loc}$ at the critical point, is the analogous of $b$.
Furthermore, the scaling of the GOE perturbation with the system size is somewhat similar to the one of the RP model~\cite{kravtsov}, a random matrix model consisting of $L$ diagonal, Poisson distributed, elements of zero mean and variance 
$\overline{({\cal H}^{\rm RP}_{ii})^2} = 1$, and $L \times L$ off-diagonal GOE matrix elements of zero mean and variance $\overline{({\cal H}^{\rm RP}_{ij})^2} = L^{- \phi}$.
The main control parameter of the problem is the exponent $\phi$: For $ 0 < \phi < 1$ the systems is fully ergodic, for $\phi > 2$ the system is fully localized, and for $1 \le \phi \le 2$ the system is in a non-ergodic extended phase.
Yet, as discussed in details in the following sections, the properties of the metal/insulator transition of the toy model considered here (as well the properties of its intermediate phase) 
are of a totally different kind with respect to both the PLRBM and the RP models, and 
do not fit the standard paradigm of Anderson localization in any dimension.

\section{The properties of the intermediate phase} \label{sec:griffiths}

In order to investigate the properties of the intermediate phase where arbitrarily large insulating and metallic regions coexist, in the following we set $\phi=1$ and consider
large values of the chain length. The intra-layer coupling is thus given by  $\gamma = \hg / L$, with $\hg$ of order $1$.

\subsection{Probability distributions of the LDoS} \label{sec:pldos}

\begin{figure}
\includegraphics[width=0.48\textwidth]{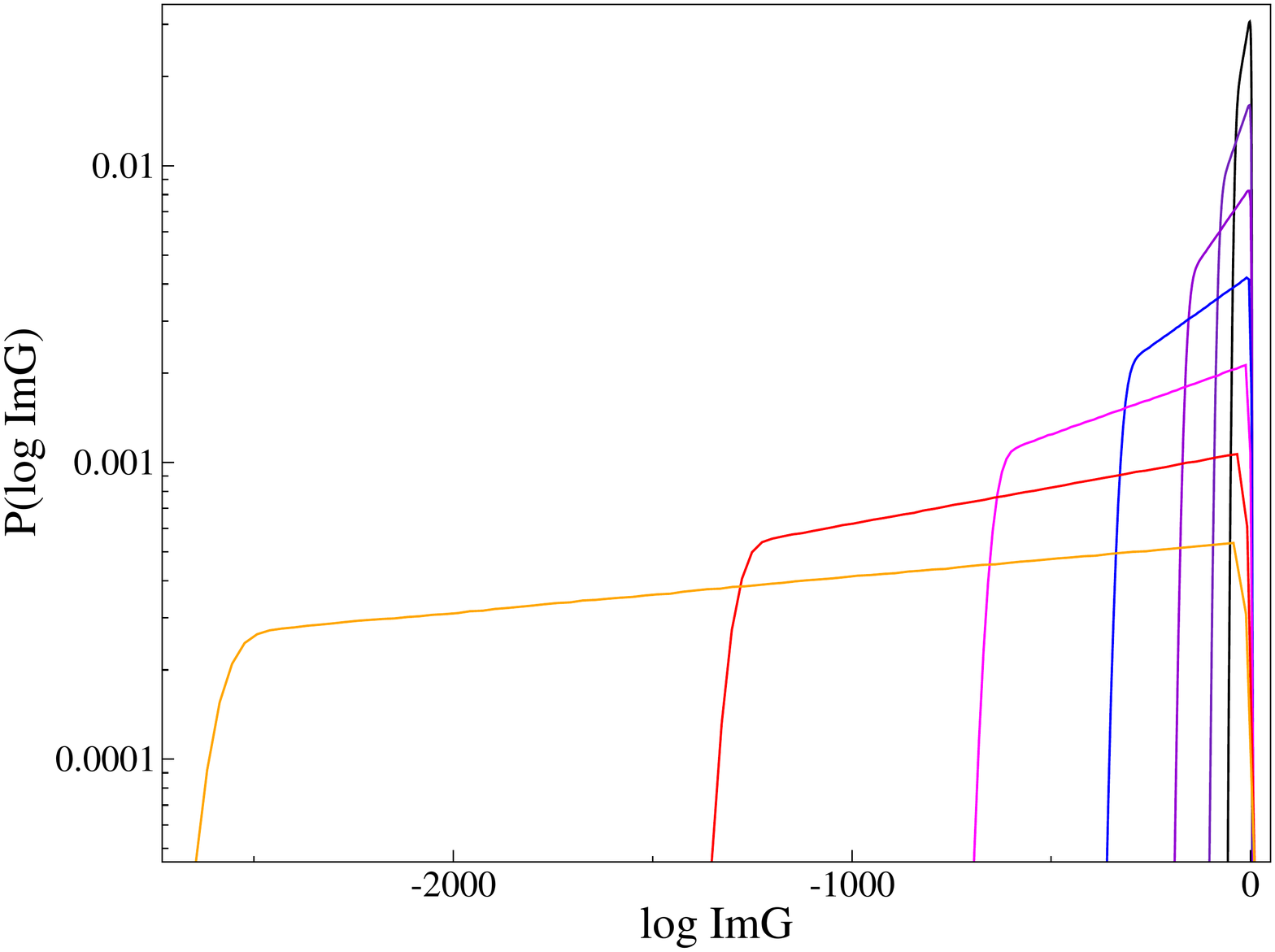}%
\caption{\label{fig:plimg}
Evolution of the probability distribution of $\log {\rm Im} {\cal G}$ (averaged over many realizations of the disorder) 
upon increasing the length of the chain for the samples that are in the conducting phase, for $\phi=1$, $W=4$, and $\hg = 2$; $L = 2^8$ (black), $2^9$ (indigo), $2^{10}$ (violet), 
$2^{11}$ (blue), $2^{12}$ (magenta), $2^{13}$ (red), and $2^{14}$ (orange).}
\end{figure}

A first important piece of information is obtained by analyzing the probability distribution of the local DoS for the samples that are in the conducting phase (i.e., the non-singular part of the distribution), shown in Fig.~\ref{fig:plimg} for $W=4$ and $\hg = 2$ and for several values of the length of the chain $L$ (a similar behavior is observed for other values of $W$ and $\hg$ in the critical region). Upon increasing the system size, $P(\log {\rm Im} {\cal G})$ tends to a flat 
distribution with a support which extends from zero to arbitrarily small values. This implies that:
\begin{equation} \label{eq:pimg}
P ({\rm Im} {\cal G}) \sim \frac{1}{{\rm Im} {\cal G}} \qquad \textrm{for~} {\rm Im} {\cal G} \in [\chi,1] \, ,
\end{equation}
(times small logarithmic corrections)
where 
$\chi$ is a $L$-dependent cut-off which goes to zero exponentially fast with $L$. This means that $\langle \log {\rm Im} {\cal G} \rangle \propto - L$, i.e.,  the typical value of the local DoS tends to zero in the thermodynamic
limit exponentially fast with $L$, while the average DoS, $\rho = \langle {\rm Im} {\cal G} \rangle / \pi$, is of order $1$: Although the system is conducting, on the vast majority of the sites of the 
LDoS can take arbitrarily small values~\cite{localized_samples}.

\subsection{Statistics of dissipation propagation and thermal inclusions} \label{sec:trans}

\begin{figure}
\includegraphics[width=0.48\textwidth]{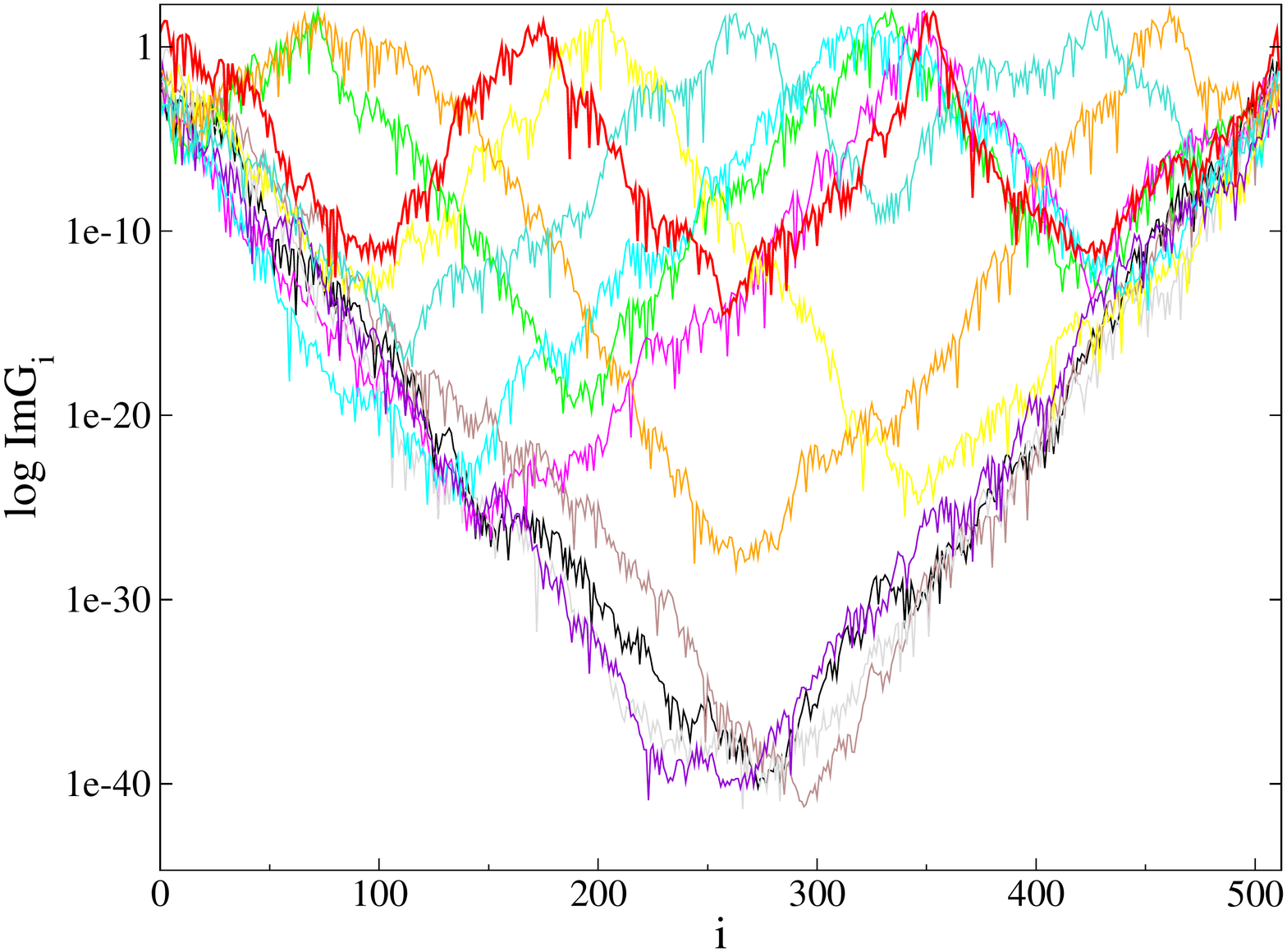}%
\caption{\label{fig:avalanches}
(Log of the) Imaginary part of ${\cal G}_{i}$ in the middle of the system when $\eta$ is set identically to zero on all the layers of the chain except on the first layer, 
where $\eta=1$, for $\phi=1$, $W=4$, $\hg=2$, and $L=2^9$. Different colors correspond to different realizations of the disorder. Some of the samples (black, gray, brown, and violet) are in the insulating phase and $\log {\rm Im} {\cal G}_i \sim - i / \xi_{\rm typ}$.
Conducting samples are characterized instead by regions where ${\rm Im} {\cal G}_i$ decays exponentially separated by few (i.e., $O(1)$) resonances (``thermal inclusions'') where
${\rm Im} {\cal G}_i$ is of order $1$.}
\end{figure}

Further insights can be obtained by studying the statistics of dissipation propagation along the chain. In order to do that we set the imaginary regulator identically equal to zero on all the positions of the chains, and put a source of dissipation on the first layer where
we set $\eta \sim 1$ (i.e., $\eta_i = \delta_{i,1}$); We solve the recursion relations~(\ref{eq:recursion}) and~(\ref{eq:recursion_final}), 
and measure the imaginary part of the Green's as a function of the position $i$.
We set $W=4$ and $\hg=2$, such that $P_{\rm loc}\sim 1/2$. In Fig.~\ref{fig:avalanches} we show the numerical results for $L=512$.
About half of the samples are in the insulating phase and ${\rm Im} {\cal G}$ drops exponentially on a characteristic scale $\xi_{\rm typ}$, until it reach a very small value 
of order $e^{-L/(2 \xi_{\rm typ})}$ in the middle 
of the chain. These samples would behave essentially in the same way if the intra-layer coupling was turned off, since the effect of $\gamma$ is only perturbatively small (i.e., 
$\xi_{\rm typ} \approx 
\xi_0$).
Conducting samples, instead, are constituted by patchworks of insulating segments, over which ${\rm Im} {\cal G}$ decays exponentially over the same length $\xi_{\rm typ}$, and few, rare
resonances (i.e., ``thermal inclusions''), where ${\rm Im} {\cal G}$ is of order $1$.
This scenario is completely different from the standard Anderson transition, and in instead consistent with the avalanche mechanism for the MBL transition put forward in Refs.~\cite{avalanches,thiery,thiery1,dimitrescu,morningstar}, in which thermalization
is driven by a network of few (i.e., $O(1)$) thermal inclusions which destabilize the insulating phase.

In App.~\ref{app:trans} we also show the whole probability distributions $\tilde{P}({\rm Im} {\cal G}_{L/2})$ of the imaginary part of the Green's at the middle of the chain, $x=L/2$ (see 
Fig.~\ref{fig:transmission}).
This analysis allows one to understand how proliferation of quantum avalanches generates a distribution of LDoS of  Fig.~\ref{fig:plimg}, which extends down to arbitrarily small values.

\subsection{Delocalization via the proliferation of quantum avalanches} \label{sec:avalanches}

\begin{figure}
\includegraphics[width=0.48\textwidth]{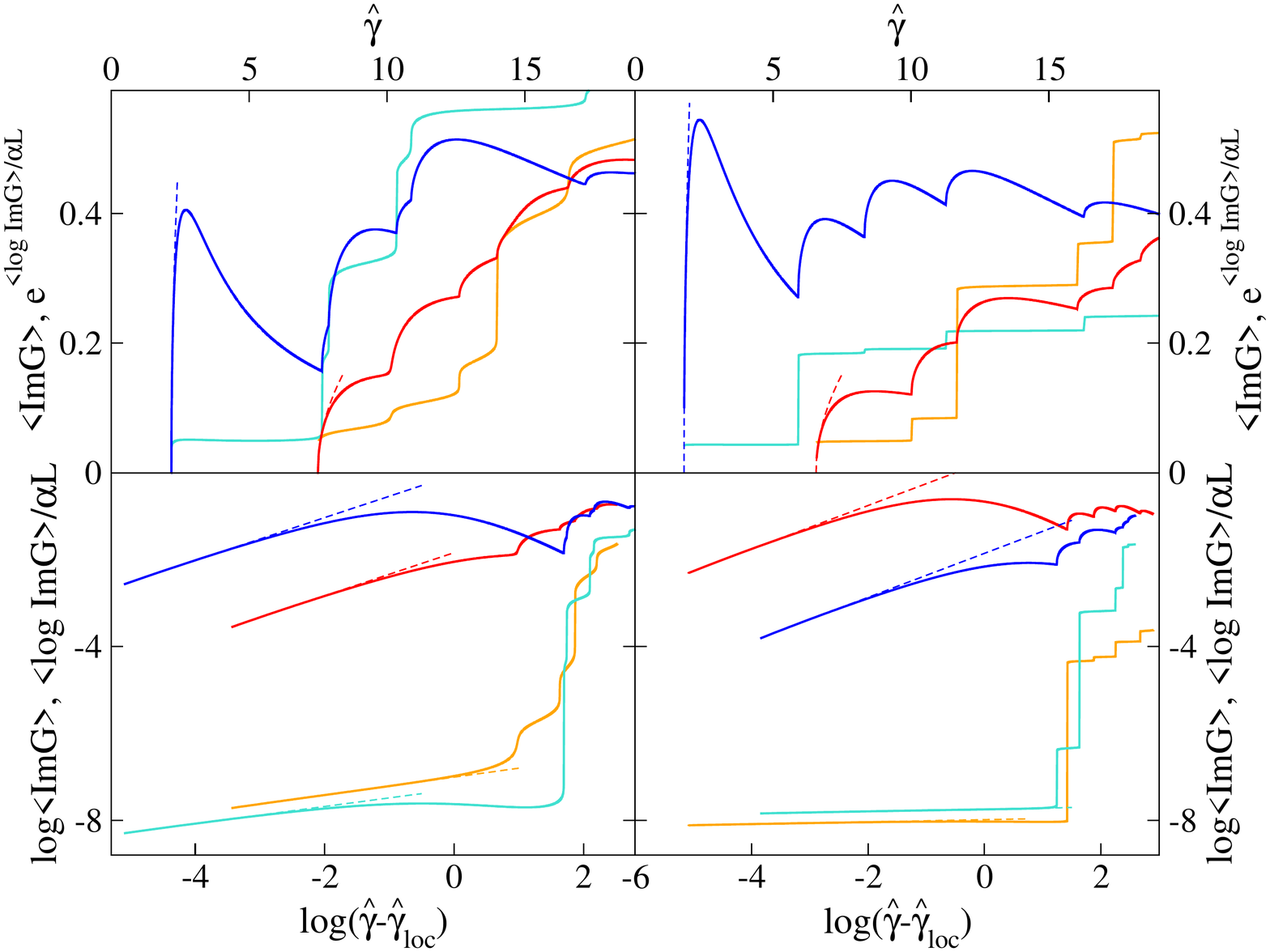}%
\caption{\label{fig:dega}
Top panels: $\langle {\rm Im} {\cal G} \rangle$ (blue and red) and $e^{\langle \log {\rm Im} {\cal G} \rangle/\alpha L}$ (light blue and orange) versus $\hat{\gamma}$ for two specific realizations of the disorder 
(with $\alpha=2.5 \cdot 10^{-2}$).
Bottom panel: $\log \langle \log {\rm Im} {\cal G} \rangle$ (blue and red) and $\langle \log {\rm Im} {\cal G} \rangle/\alpha L$ (light blue and orange) 
versus $\log(\hat{\gamma} - \hat{\gamma}_{\rm loc})$ for the same
samples (with $\alpha \approx 10^{-2}$). 
The size of the chain is $L=2^8$ (left panels) and $L=2^{11}$ (right panels) and $W=4$, $\phi=1$.
The dashed lines correspond to the power-law fits of the critical behavior of Eqs.~(\ref{eq:critical}).}
\end{figure}

The unconventional mechanism for delocalization of the model is clearly illustrated by Fig.~\ref{fig:dega}, which shows how the insulating phase is destabilized by the proliferation of quantum avalanches when the coupling with the GOE-like perturbation is increased.
In the figure we plot the behavior of
$\langle {\rm Im} {\cal G} \rangle$ 
and $\langle \log {\rm Im} {\cal G} \rangle$ 
as a function of $\hat{\gamma} = L \gamma$ for two specific samples and two system sizes ($L=2^8$, left panels, and $L=2^{11}$, right panels), when the imaginary regulator is set identically equal
to $\eta = 0^+$ on all sites of the chain. Here we denote with angular brackets $\langle\ldots\rangle$ the real-space average over different sites, at fixed disorder realization. If $\hat{\gamma}$ is smaller than a critical threshold $\hat{\gamma}_{\rm loc}$ (which depends on the disorder realization and is exponentially distributed~\cite{pgammaloc}), 
the system is a Anderson insulator and $\langle {\rm Im} {\cal G} \rangle = 0$.
At the threshold $\hat{\gamma}_{\rm loc}$ the first resonance is formed due to the GOE-like intra-layer coupling, and the Green's functions spontaneously develops a non-vanishing imaginary part, corresponding to the fact that dissipation starts to propagate throughout the chain.
The critical behavior at the transition is given by:
\begin{equation} \label{eq:critical}
\begin{aligned}
& \langle {\rm Im} {\cal G} \rangle \sim a \, ( \hat{\gamma} - \hat{\gamma}_{\rm loc})^\nu \, , \\
& e^{\langle \log {\rm Im} {\cal G} \rangle} \sim b_L ( \hat{\gamma} - \hat{\gamma}_{\rm loc})^\nu \, ,
\end{aligned}
\end{equation}
with $\nu = 0.5$ (dashed lines). The constant $a$ is of order $1$, while $b_L$ is exponentially small in the system size, $b(L) \sim e^{- \alpha L}$ 
(with $\alpha \propto \xi_{\rm typ}^{-1}$).
This is due to the fact that the average DoS 
is domniated by the resonances, i.e., the extreme values of the distribution $P({\rm Im} {\cal G})$, while each 
single resonance only yields a contribution of order $e^{-L/\xi_{\rm typ}}$ to the typical value of the LDoS (see Figs.~\ref{fig:plimg} and~\ref{fig:avalanches}).
When $\hat{\gamma}$ is further increased beyond $\hat{\gamma}_{\rm loc}$, more resonances are formed. Each new resonance produces a cusps in the average DoS and 
an sharp increase in the typical value of the LDoS~\cite{cusp}, i.e., an avalanche.
In fact, when a new resonance appears, the typical LDoS increases by a factor 
proportional to the distance between the new resonance and the closest pre-existing one, which
is essentially the portion of the chain that has become ``thermal'' due to the appearence of the new resonance. 
A similar behavior is found upon increasing the disorder $W$ at fixed $\hat{\gamma}$. The statistics of the avalanche size distribution is extensively discussed in App.~\ref{app:avalanches}.
Note that the separation between the formation of two successive resonances is of $O(1)$ in $\hat{\gamma}$, independently of $L$. Hence in the thermodynamic limit an infinite number (i.e. of order $L$) of resonances appears when one goes from $\phi \to 1^+$ to $\phi \to 1^-$. Each new resonance produces a singularity in the typical value of the LDoS of the kind of the one described by Eq.~(\ref{eq:critical}). We argue that the condensation in a single point of infinitely many square root singularities with an exponentially small prefactor  can lead to a much sharper non-analyticity of the typical value of the LDoS at $\phi=1$, and possibly yield an exponential critical behavior as the one expected for a KT-like criticality~\cite{goremykina,dimitrescu,morningstar}. 
Within this interpretation the proliferation of quantum avalanche (although they are not topological excitations) intuitively resembles vortex unbinding at the KT transition~\cite{dimitrescu}.

The anomalous behavior of the intermediate phase 
is also confirmed by the study of the probability distribution $L(\lambda)$ of the local Lyapunov exponents (see App.~\ref{app:lyap}), which describe 
the exponential growth or the exponential decrease of the imaginary part of the Green's functions with the number of recursion steps 
when one starts from an infinitesimally small value. We find that the distributions $L(\lambda)$ exhibit exponential tails. This behavior is consistent with that of a quantum Griffiths phase~\cite{Vojta} characterized by
exponential rare large insulating regions with exponentially large resistance.

\subsection{Fractal thermal inclusions 
and Kosterlitz-Thouless Criticality}
\label{sec:fractal}

\begin{figure}
\includegraphics[width=0.48\textwidth]{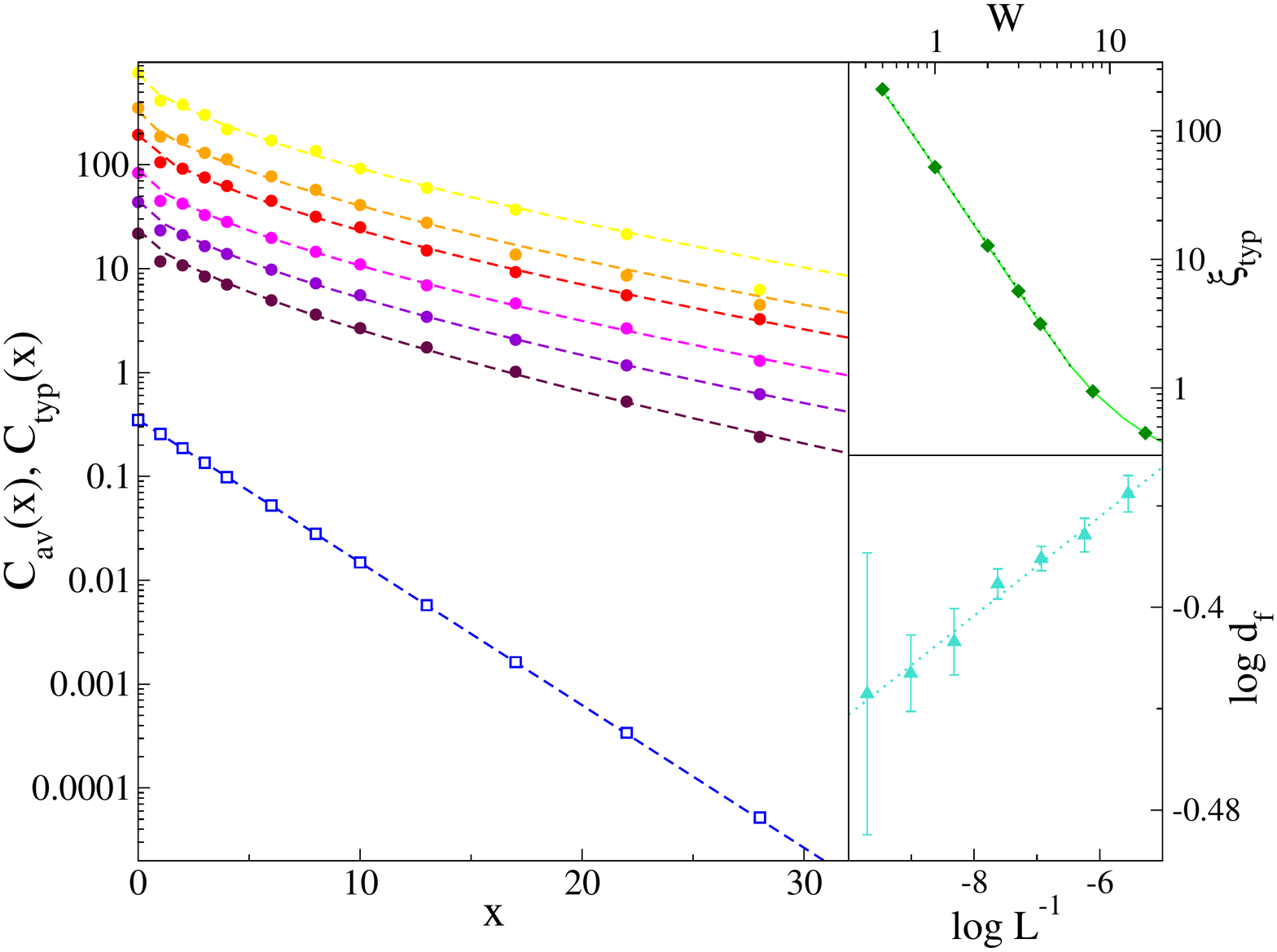}%
\caption{\label{fig:corr}
Main panel: Average (filled circles) and typical (blue empty squares) correlation functions for $\phi=1$, $W=4$ and $\hg=2$ (a similar behavior is observed for all values of $\hg$ and $W$ at the critical point). 
Typical correlations do not depend on the length of the chain, whereas average correlations increase as $L$ is increased ($L=2^8$, maroon, $L=2^9$, violet, $L=2^{10}$, magenta, $L=2^{11}$, red, $L=2^{12}$, orange, $L=2^{13}$, yellow)
The dashed lines correspond to fits of the exponential and sub-exponential decays of $C_{\rm typ}(x)$ and $C_{\rm av}(x)$, Eqs.~(\ref{eq:corr}), with $\xi_{\rm typ} \approx 3.16$.
Top inset: Localization length of typical correlation functions, $\xi_{\rm typ}$, as a function of $W$.  
The green dotted line corresponds to a fit as $\xi_{\rm typ} = A/W^2$, with $A \approx 52.3$. Bottom inset: 
(Log of the) Fractal exponent $d_{\rm f}$ extracted from the stretched exponential fits of average correlations as a function of the (log of the) inverse length of the chain. The dotted line corresponds
to a power-law fit as $d_{\rm f} \sim L^{-\kappa}$, with $\kappa \approx 0.02$.}
\end{figure}

Another very important (and tightly related) feature of the model in the critical region is represented by the fractal behavior of correlation functions between points at distance $x$ along the chain.
The probability that a ``particle'' starting on a certain site $(i,p)$ at time $t=0$ (i.e., $\vert \psi (t=0) \rangle = \vert i,p \rangle$) is found at distance $x$ along the horizontal direction 
in the long time limit has a simple spectral representation as:
\[
\lim_{t \to \infty} \vert \langle i + x,p \vert e^{- i H t / \hbar} \vert i,p \rangle \vert^2 \propto \vert {\cal G}_{i,i+x} \vert^2 \, ,
\]
where ${\cal G}_{i,i+x}$ is the off-diagonal element on sites $i$ and $i+x$. Such off-diagonal element can be easily expressed in terms of the diagonal elements of the Green's functions
and of the cavity Green's functions only as:
\[
\begin{aligned}
{\cal G}_{i,i+x} = & t G^{(r)}_i \, t G^{(r)}_{i+1} \cdots t G^{(r)}_{i+x-1} \, {\cal G}_{i+x} \\
= & t G^{(l)}_{i+x} \, t G^{(l)}_{i+x-1} \cdots t G^{(l)}_{i+1} \, {\cal G}_{i} \, . \\
\end{aligned}
\]
Since correlations are likely dominated by rare realizations of the disorder and/or rare insulating or very weakly conducting segments of the chain, 
it is useful to measure both its average $C_{\rm av}(x)$ and typical $C_{\rm typ}(x)$ values. These quantities are plotted in Fig.~\ref{fig:corr} as a function of the distance $x$ for $W=4$ and $\hg=2$ (the same features are observed at other values of $\hg$ and $W$ at criticality), showing 
an apparent different behavior:
\begin{equation}
\begin{aligned} \label{eq:corr}
C_{\rm av} (x) &=\overline{ \langle | {\cal G}_{i,i+x} |^2 \rangle}  \approx A(L) \, e^{-(x/\xi_{\rm typ})^{d_{\rm f}(L)}} \, , \\
C_{\rm typ} (x) & = e^{\overline{\langle \log | {\cal G}_{i,i+x} |^2 \rangle}} \approx B \, e^{-x/\xi_{\rm typ}} \, .
\end{aligned}
\end{equation}
We note that the overline here denotes averaging over disorder realizations, while brackets indicate average over sites.
While $C_{\rm typ} (x)$ shows the usual exponential decay with distance $x$ over the characteristic length $\xi_{\rm typ}$, $C_{\rm av}(x)$ decay as stretched exponentials with an exponent $d_{\rm f} (L)$.
Furthermore, typical correlations do not depend on the length of the chain (i.e., the prefactor $B$ is a constant independent of $L$), whereas average correlations increase as $L$ is increased (for $W=4$ we find that $A(L) \sim L^{\psi}$ with 
$\psi \approx 0.18$ independently of $\hg$). 
These features have been already highlighted in Refs.~\cite{zhang,thiery,goremykina,morningstar}, and can be interpreted in terms of the fractal structure of thermal inclusions 
(i.e., a fractal set of rare locally thermalizing regions).
We find that the fractal exponent $d_{\rm f}$ extracted from the fits of numerical data slowly but systematically decreases with the system size (bottom right panel of Fig.~\ref{fig:corr})
from $d_{\rm f} \approx 0.701$ for $L=2^8$ to $d_{\rm f} \approx 0.648$ for $L=2^{14}$.

As shown in the top inset of Fig.~\ref{fig:corr}, the localization length of typical samples, $\xi_{\rm typ}$, grows as the disorder is decreased and diverges as $\xi_{\rm typ} \sim W^{-2}$, proportionally to the localization length $\xi_0$ of 
the unperturbed case ($\gamma = 0$).
Thus, for two points at a given distance $x$ on the chain, typically the effect of the intra-layer coupling $\gamma$ is just to increase the localization length perturbatively by a small factor compared to the $\gamma=0$ limit.

\begin{figure}
\includegraphics[width=0.48\textwidth]{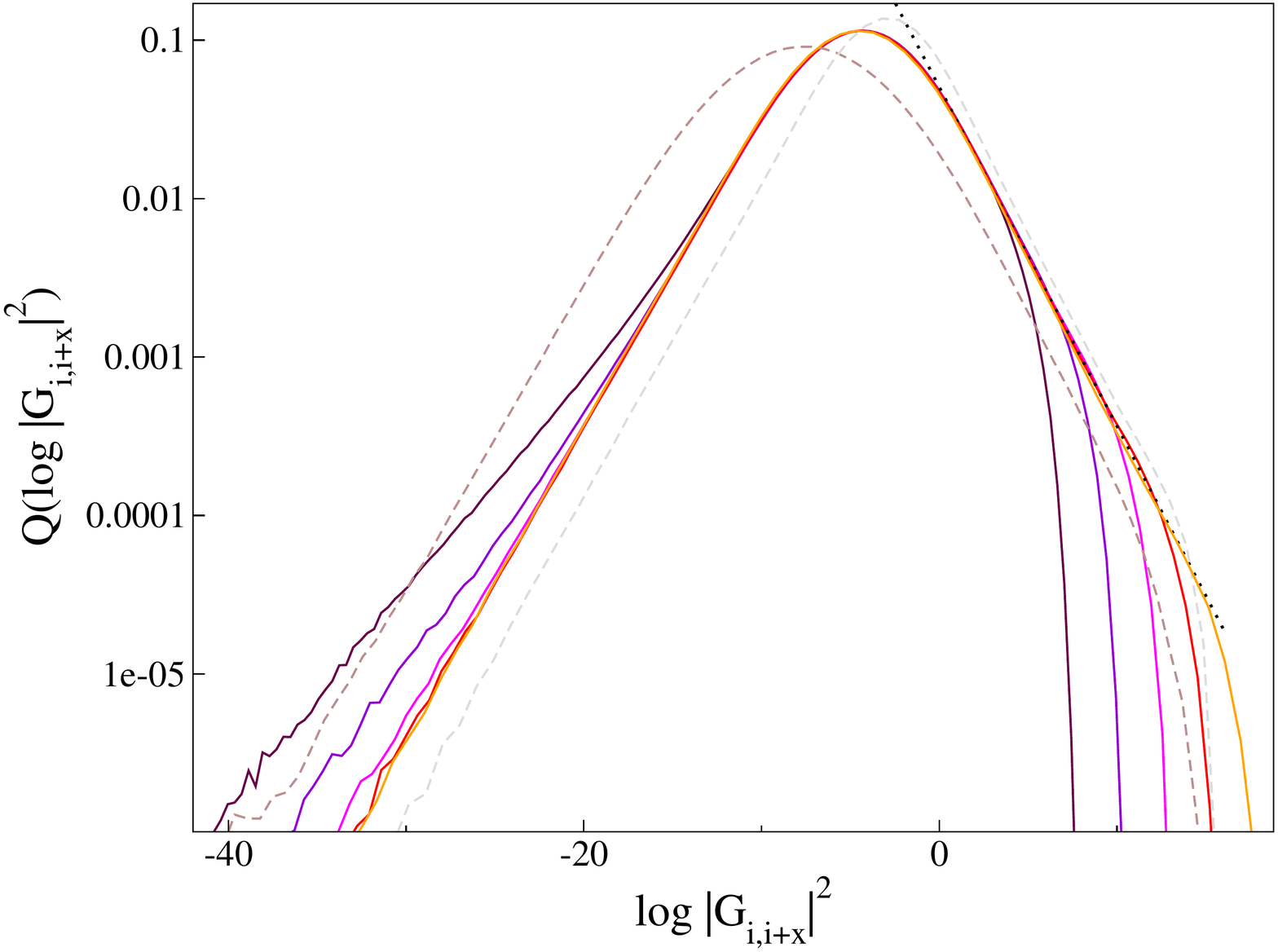}%
\caption{\label{fig:pcorr}
Probability distributions $Q ( \log | {\cal G}_{i,i+x} |^2)$ at fixed distance $x$ for $\phi=1$, $W=4$, and $\hg=2$. Continuous curves corresponds to the pdf's for $x = 10 \approx 3 \xi_{\rm typ}$ varying the system size
($L=2^8$, maroon, $L=2^{10}$, violet, $L=2^{12}$, magenta, $L=2^{14}$, red, $L=2^{16}$, orange). 
The black dotted line is the power-law fit of the tails as $Q (| {\cal G}_{i,i+x} |^2) \sim [| {\cal G}_{i,i+x} |^2]^{-3/2}$. The dashed curves show the pdf's
varying the distance ($x \approx 6.4 \xi_{\rm typ}$, brown, and $x \approx 1.6 \xi_{\rm typ}$, gray) for the largest system size $L=2^{14}$.}
\end{figure}

An equivalent way to interpret the fact that average correlations grow with the system size and decay much slowlyer than typical correlations 
can be achieved by realizing that the matrix elements $| {\cal G}_{i,i+x} |^2$ at fixed distance $x$ are in fact broadly distributed.
The probability distribution $Q ( \log | {\cal G}_{i,i+x} |^2)$ is plotted in Fig.~\ref{fig:pcorr} for $W=4$ and $\hg=2$ (a similar behavior is observed varying $W$ and $\hg$). The continuous curves correspond to the 
distribution functions for $x \approx 3 \xi_{\rm typ}$ varying the length of the chain. 
The black dotted line correspond to a fit of the tails of the pdf as:
\[
Q (| {\cal G}_{i,i+x} |^2) \sim \frac{1}{\left[| {\cal G}_{i,i+x} |^2\right]^{3/2}} \, , \,\,\,\,\, \textrm{for~} | {\cal G}_{i,i+x} |^2 \in [a,\Lambda] \, .
\]
The lower cut-off $a$ is of the order of the typical value and does not depend on $L$, while the upper cut-off $\Lambda$ increases with the system size as $\Lambda \sim L^\theta$.
These plots show that at fixed $x$, the typical value of $| {\cal G}_{i,i+ 3 \xi_{\rm typ}} |^2$, 
i.e., $C_{\rm typ} (3 \xi_{\rm typ})$, is independent of $L$ and finite. Conversely, the average value of $| {\cal G}_{i,i+3 \xi_{\rm typ}} |^2$, i.e., $C_{\rm av} (3 \xi_{\rm typ})$ 
is dominated by the fat tails of the distributions and diverges as $\sqrt{\Lambda} \sim L^{\theta/2}$ in the thermodynamic limit. (Since we have 
that $A(L) \sim L^\psi$, this implies that $\theta = 2 \psi$.)
The figure also shows the probability distributions $Q ( \log | {\cal G}_{i,i+x} |^2)$ at $x \approx 6.4 \, \xi_{\rm typ}$ (dashed brown) and $x \approx 1.6 \, \xi_{\rm typ}$ (dashed gray) for the largest system size $L=2^{14}$.
The only effect of varying the distance $x$ is to shift the whole distributions (and thus the typical value) to the left or to the right, without otherwise modifying their qualitative behavior.

This analysis indicates that 
for most of the pair of sites at distance $x$ of a given sample, correlations are equal to $e^{-x/\xi_{\rm typ}}$, with $\xi_{\rm typ}$ proportional to $\xi_0$ and finite. 
Yet, there are few, rare 
positions 
for which correlations can be much larger. 
In other words, the localization length $\xi_{\rm typ}$ of typical segments is finite at the critical point 
and it is just proportional to the one of the unperturbed limit ($\gamma=0$), while the average correlation length diverges
in the thermodynamic limit due to the presence of
rare thermal inclusions where the localization length can become arbitrarily large (i.e., of the order of the system size). In particular, assuming a distribution of localization lengths $\Xi(\xi)$ and assuming 
that $| {\cal G}_{i,i+x} |^2 = e^{-x/\xi}$, 
one obtains that $\xi$ is also broadly distributed, with typical value $\xi_{\rm typ}$ and power-law tails 
\begin{equation} \label{eq:Pxi}
\Xi(\xi) \sim x/\xi^2 
\end{equation}
which dominate the average.

It is worth emphasizing the similarity between the results highlighted here and those obtained from the numerical solution of the RGs schemes for the MBL transition of Refs.~\cite{zhang,morningstar,goremykina,dimitrescu}.
From one side, the stretched exponential behavior of the average correlation in Eq.~(\ref{eq:corr}) leads to a fractal dimension $d_f(L)$ 
which is numerically very close to the value $d_f = \log 2/ \log 3 = 0.631$ found in the toy RG solved in~\cite{zhang} for the range of system sizes explored.
%
While the latter RG scheme assumed a symmetry between the thermal and the MBL phase (and obeys one-parameter scaling), later modifications of this toy RG without such a constraint have been developed, leading to a two parameter KT-like RG flow~\cite{goremykina,morningstar}.  Subsequent work~\cite{dimitrescu} argued that in fact KT-like RG flow follows generally from considering an MBL transition driven by avalanches~\cite{avalanches,thiery1,thiery}.
In this respect it is worth emphasizing that our toy model~(\ref{eq:H}) lacks this symmetry  (few, rare resonances can destabilize the insulating phase but few rare insulating regions cannot prevent 
the spreading of the wave-packet) and its delocalization transition appears to be driven by quantum avalanches. This suggests that also our delocalization transition could infact belong to the same universality class, and that $d_{\rm f}$ tends to $0$ for $L \to \infty$. However, as also recently shown in~\cite{morningstar}, determining numerically the asymptotic KT critical behavior from the analysis of finite size samples is a very hard task and $d_{\rm f}$ is affected by strong finite-size effects. Finally, we notice that  the broad distribution of the localization length of Eq.~(\ref{eq:Pxi}) is also in agreement with results from RG approaches~\cite{goremykina,morningstar,thiery} and with a recent numerical study~\cite{herviou} of the MBL transition, and it is consistent with the KT-type criticality~\cite{dimitrescu,morningstar}. 

\section{Slow dynamics and anomalous transport} \label{sec:dynamics}

In this section we focus on the implications of the coexistence of arbitrarily large conducting and insulating segments on the dynamical and transport properties of the system.
We focus on several observables that have been used to probe the unusual behavior that emerges in the bad metal delocalized phase preceding the many-body localization
transition~\cite{dave1,demler,BarLev,alet,doggen,evers,luitz_barlev}, such as the large time behavior of the return probability, the mean square displacement, and the low-frequency behavior of the 
optical conductivity. The first two observables are local probes, while the latter probes the long-wavelength behavior of the system.

Our initial state $\vert \psi_0 \rangle$ correspond to a ``particle'' sitting on a site $\vert i, p \rangle$, with $p$ randomly chosen among the $M$ sites belonging to the layer $i$ with energy $\epsilon_i$ close to $0$ (i.e., in the middle of the spectrum, corresponding to high temperature).
The wave function at time $t$ (we rescaled the time by $1/\hbar$) can be written in terms of the eigenvalues $E_\alpha$ and the eigenfunctions $\alpha$ of the Hamiltonian~(\ref{eq:H})
as
\[
\vert \psi (t) \rangle = \sum_\alpha e^{-i E_\alpha t} \langle \alpha \vert i , p \rangle \vert \alpha \rangle \, .
\]
The return probability $R(t)$ is defined as the probability to find the ``particle'' on the $i$-th layer after time $t$:
\[
R(t) = 
\overline{\left \langle \sum_{q=1}^M \left \vert \sum_\alpha e^{-i E_\alpha t} \langle \alpha \vert i , p \rangle \langle i, q \vert \alpha \rangle \right \vert^2 \right \rangle} \, ,
\]
where the average is performed over several starting layers $i$ with $\epsilon_i$ close to zero energy, and over the disorder distribution.
Analogously, one can define the mean-square displacement as the square of the average distance along the $x$ direction traveled by the ``particle'' after time $t$:
\[
\left \langle x^2 (t) \right \rangle =
\overline{ \left \langle 
\sum_{r=-L/2}^{L/2} \! 
r^2 \sum_{q=1}^M \left \vert \sum_\alpha e^{-i E_\alpha t} \langle \alpha \vert i , p \rangle \langle i+r,q \vert \alpha \rangle \right \vert^2 \right \rangle }\, .
\]
We have computed these two dynamical observables by exact diagonalizations of finite size samples. We have varied $L$ from $24$ to $48$ and taken 
$M = q N$, with the ratio $q$ ranging from $4$ to $8$, finding no significant dependence on the different values of $L$ and $M$ chosen~\cite{largeM}. Numerical data are
averaged over $64$ independent realizations of the disorder.

\begin{figure}
\includegraphics[width=0.48\textwidth]{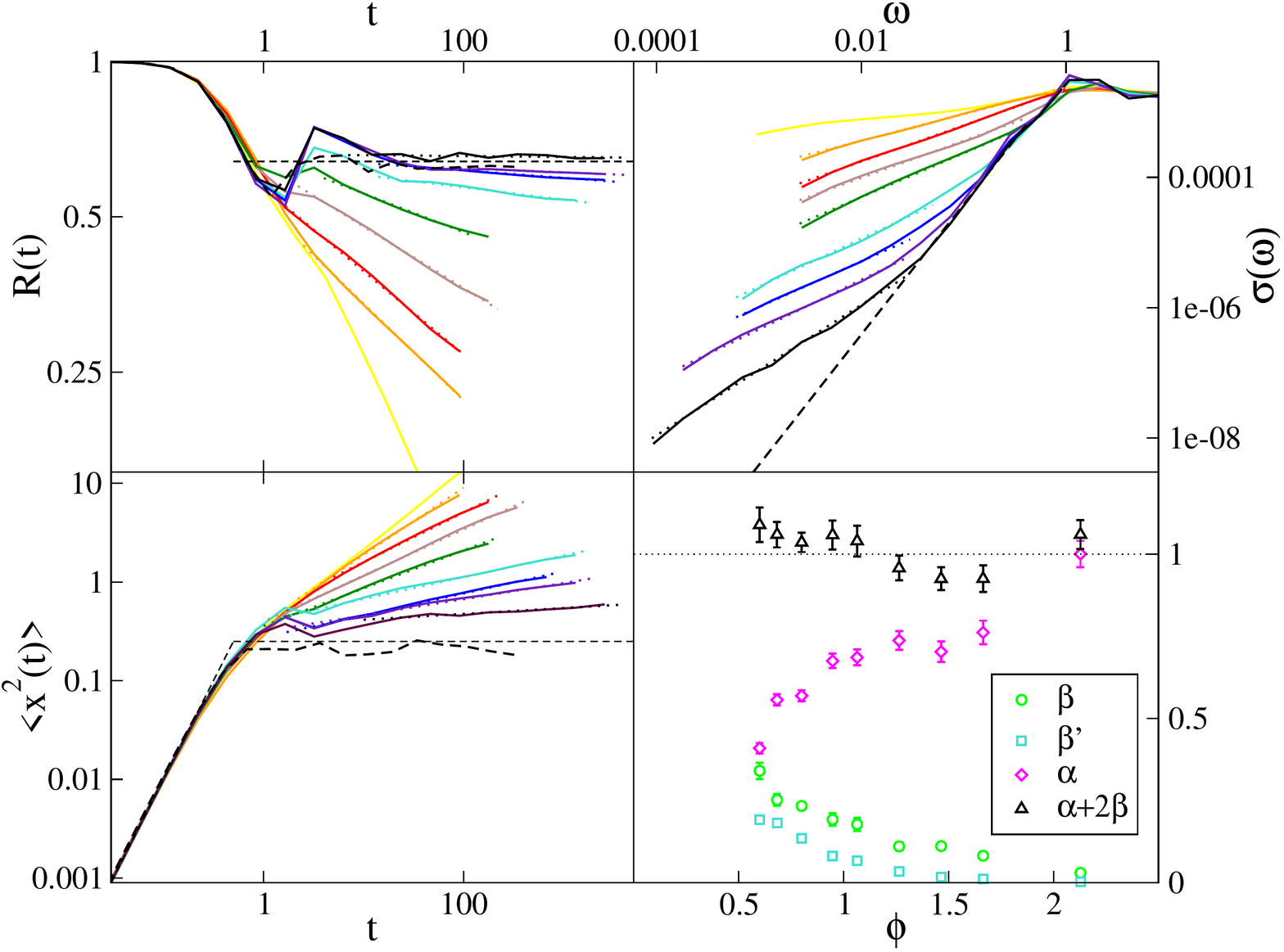}%
\caption{\label{fig:dynamics}
Top left panel: Return probability as a function of time for $W=16$, $L=36$, and $M=160$, and for several values of $\gamma$ across the
intermediate region (corresponding to the light blue squares  in the interval $\phi \in [0.5,2.2]$ on the phase diagram of Fig.~\ref{fig:phase_diagram}). The black dashed curve shows the return probability for the standard $1d$ Anderson model ($\gamma=0$) and the horizontal dotted black line 
gives the value of the IPR of the unperturbed Anderson localized wave-functions. Bottom left panel: Mean square displacement as a function of $t$ for the same values of $W$, $L$, $M$, and $\gamma$. The black dashed curve shows $\langle x^2 (t) \rangle$ for the standard $1d$ Anderson model ($\gamma=0$) and the horizontal dotted black line gives the square of the 
bare localization length $\xi_0$ in $1d$. Top right panel: Infinite-temperature optical conductivity $\sigma (\omega)$ for the same values of $W$, $L$, $M$, and $\gamma$.
A factor $T$ is implicitly understood.
The black dashed curve correspond to the standard $1d$ Anderson model ($\gamma=0$), for which we recover $\sigma(\omega) \sim \omega^2 \log^2 (\omega)$ (in $1d$).
Bottom right panel: Numerical values of the exponents $\beta$, $\beta^\prime$, and $\alpha$, obtained by power-law fits of the numerical data, as a function of $\phi$.
The scaling relation $\alpha + 2 \beta = 1$ is roughly satisfied within our numerical accuracy (black diamonds), while we find that $\beta^\prime \approx 0.6 \beta$~\cite{evers}.}
\end{figure}

The results are plotted in the top and bottom left panels of Fig.~\ref{fig:dynamics} for $W=16$ (such that $\xi_0 \ll L$) and several values of $\phi$ across the intermediate Griffiths region.
The return probability and the mean square displacement display slow dynamics and power laws strikingly similar to those observed in recent simulations and experiments 
in the bad metal delocalized phase preceding MBL~\cite{dave1,BarLev,luitz_barlev,demler,alet,doggen,evers}.
On short time scales, i.e., $t$ of order $1$, the system behaves as the standard Anderson insulator in $1d$ ($\gamma=0$) for any values of $\phi$: 
The mean square displacement grows ballistically $\langle x^2 (t) \rangle \sim t^2$ until the wave-packet spreads over the bare localization length $\xi_0$ of the $1d$ disordered 
tight-binding model in absence of the intra-layer coupling, and $R(t)$ decays exponentially
to the Inverse Participation Ratio (which is of order of $1/\xi_0$) of the unperturbed Anderson localized eigenstates close to the middle of the band.
The effect of the GOE coupling sets in on larger time scales, $\xi_0 < t < 1/\delta$ (with $\delta \sim 1/(LM)$ being the mean level spacing), where a regime of slow and anomalous dynamics emerges and both observables show a clear 
power-law behavior:
The mean square displacement grows sub-diffusively as $\langle x^2 (t) \rangle \sim t^{2 \beta}$ and the return probability decays as $R(t) \sim t^{-\beta^\prime}$, with exponents 
that decrease smoothly as $\phi$ is increased (i.e., $\gamma$ is decreased). 
At even larger times, asymptotic time dynamics is determined by finite size effects such as reflections from the boundaries.

In the diffusive regime ($t \gg \xi_0$) one naturally expects that
\[
R(t) \propto \frac{1}{\sqrt{\left \langle x^2 (t) \right \rangle}} \, ,
\]
and hence $\beta = \beta^\prime$.
Numerically, we find that $\beta^\prime$ is smaller than $\beta$ by a factor approximately equal to $0.6$ (see bottom right panel of Fig.~\ref{fig:dynamics}). 
This discrepancy might be either due to due to finite 
size effects 
or (more likely) to the fact that the spreading of the wave-packet in time is not described by a Gaussian shape, as recently reported in Ref.~\cite{evers}.

Using exact diagonalizations we also examine the infinite-temperature low-frequancy behavior of the optical conductivity along the $x$ direction, $\sigma (\omega)$.
Using linear response and the Lehmann representation of $T \sigma (\omega)$, the real part of the conductivity in the infinite $T$ limit reads:
\[
T \sigma (\omega) = 
\overline{\frac{1}{L Z} \sum_{\alpha,\beta} \left \vert \left \langle \alpha \right \vert \hat{J} \left \vert \beta \right \rangle \right \vert^2 \delta (\omega
- E_\alpha + E_\beta )} \, ,
\]
where the current operator $\hat{J}$ is related to the creation and annihilation operators $d^\dagger_{i,p}$ and $d_{i,p}$ from the continuity equation along the $x$ direction:
\[
\hat J = i t \sum_{i=1}^L \sum_{p=1}^M \left ( d^\dagger_{i,p} d_{i+1,p} - d^\dagger_{i+1,p} d_{i,p} \right) \, .
\]
The numerical results, plotted in the top right panel of Fig.~\ref{fig:dynamics}, indicate that at low frequency $\sigma( \omega) \sim \omega^\alpha$.
The anomalous power-law regime sets in at lower and lower frequencies as $\gamma$ is decreased, and in the $\gamma \to 0$ limit one recover the expectation for the 
standard noninteracting Anderson insulator $\sigma(\omega) \sim \omega^2 \log^2 (\omega)$.
Since $T \sigma (\omega) \sim D (\omega)$, where $D(\omega)$ is the Fourier transform of the effective diffusion coefficient $D(t) = \langle x^2 (t) \rangle /t \sim t^{2 \beta - 1}$,
one expects that $\alpha + 2 \beta = 1$. In the top right panel of Fig.~\ref{fig:dynamics} we plot the values of the exponents $\beta$, $\beta^\prime$, and $\alpha$ obtained
by power-law fits of the numerical data, and show that the scaling relation is well satisfied within our numerical precision.

Furthermore, we have computed the full distribution $R[\rho(\omega)]$ of resistivities
$\rho=1/\sigma$ at a fixed sample size as a function of frequency; 
We find that the distribution of resistivities grows increasingly broad at low frequencies.
In particular, in the low frequency limit the distribution approaches a power-law $R(\rho) \sim \rho^{-\tau}$ (see Fig.~\ref{fig:psigma} of App.~\ref{app:resist}). 
We find that $\tau$ slightly increases
from $\tau \approx 1.3$ to $\tau \approx 1.6$ when increasing $\phi$, although we are unable to reliably extract the
exponent $\tau$ directly from the data, owing to the difficulty of taking the dc limit in a finite system.
The power-law tails of $R(\rho)$ imply  that the width (and sufficiently high moments of the resistivity distribution) diverges for $\omega \to 0$~\cite{demler}.
Such behavior is characteristic of a quantum Griffiths phase~\cite{Vojta}, in which
power-law correlations emerge due to the interplay between the exponential rareness of large insulating regions and their exponentially large resistance.

\section{Effective one dimensional model} \label{sec:1deff}

The toy model we have discussed so far can be seen as a $(1+\infty)$-dimensional generalization of the Anderson model. Yet it features a number of properties which are remarkably different from standard Anderson Localization in any finite dimension and, as we have been trying to argue, shares quite some similarities with the known phenomenology of one dimensional MBL systems. Since this is particularly true for what concerns properties along the longitudinal one-dimensional-like direction, it would be tempting to effectively eliminate the orthogonal \emph{infinite-dimensional} GOE perturbation described by the intra-layer coupling $\gamma$ and obtain an effective one-dimensional model.

In appendix~\ref{app:eliminating} we show how this can be achieved from the recursion equations. In particular, by eliminating the cavity Green's function $G_i^{(v)}$, we obtain a closed effective one dimensional recursion for the left-right cavity Green's functions $G_i^{(l)},G_i^{(r)}$. This recursion, differently from the one of the non-interacting Anderson problem becomes non-linear and couples together left and right cavity Green's functions, due to the presence of a self-energy correction~(\ref{eq:recursion_sigma}) which mimics the effect of a many body interaction.

A different perspective can be obtained by going back to the recursive equations~(\ref{eq:recursion}) and~(\ref{eq:recursion_final})  and noticing that those can be in fact interpreted as the recursive equations for an effective Anderson tight-binding model on a $1d$ chain of length $L$ in presence of modified on-site energies:
\begin{equation} \label{eq:Heff}
\begin{aligned}
{\cal H}_{1d}^{\rm eff} =& - \sum_{i=1}^L \left[ \tilde{\epsilon}_i \, d^\dagger_{i} d_{i} 
+ t 
\left ( d^\dagger_{i} d_{i+1} + \textrm{h.c.} \right) \right] \, ,
\end{aligned}
\end{equation}
with 
\[
\tilde{\epsilon}_i = \epsilon_i + (k+1) \gamma^2 G_i^{(v)} \, ,
\]
where the $G_i^{(v)}$'s should be self-consistently determined from Eq.~(\ref{eq:recursion}).
The modified on-site energies are complex and strongly correlated~\cite{1dcorrelated}, as they depend on all the other $\epsilon_i^\star$'s via 
Eq.~(\ref{eq:recursion}). In particular, from the last of Eqs.~(\ref{eq:recursion}) one has that:
\[
\frac{(k+1) \gamma^2}{\tilde{\epsilon}_i - \epsilon_i} = - \frac{k \tilde{\epsilon}_i + \epsilon_i}{k+1}  - z - t^2 G_{i+1}^{(l)} - t^2 G_{i-1}^{(r)}  \, .
\]
Neglecting the correlations between the $\epsilon_i$'s and the $G_j^{(v)}$'s, 
one has that:
\[
 \left \langle \tilde{\epsilon}_i \, \tilde{\epsilon}_{i+x}^{\, \star} \right \rangle \approx (k+1)^2 \gamma^4 \left \langle 
 G_i^{(v)} \left( G_{i+x}^{(v)} \right)^{\! \star} \right \rangle \, ,
 \]
One can show~\cite{tikhonov} that the correlation function $\langle G_i^{(v)} (G_{i+x}^{(v)})^\star \rangle$ controlling the spatial correlations of the modified on-site energies $\tilde{\epsilon}_i$'s of the effective $1d$ Anderson Hamiltonian is directly related to the correlation function  $\langle | {\cal G}_{i,i+x} |^2 \rangle$  studied in section~\ref{sec:fractal}. 
 At the critical point they are broadly distributed, with typical values decaying exponentially with the 
 distance as $e^{-x/\xi_{\rm typ}}$ (with $\xi_{\rm typ}$ proportional to $\xi_0$ and finite), and power-law tails decaying with an
 exponent $3/2$ (see Sec.~\ref{sec:fractal} and Figs.~\ref{fig:corr} and~\ref{fig:pcorr} for more details).
 The correlation between the modified random energies at distance $x$ is thus typically short range, but there are rare pair of sites where correlations can
 be arbitrarily strong.
 Furthermore, the fact that the $\tilde{\epsilon}_i$ are complex indicates that the GOE coupling acts locally as a thermal bath by providing 
 an effective dissipation, as it will 
 be discussed further in App.~\ref{app:self}.

\section{Conclusions and Perspectives} \label{sec:conclusions}

In this paper we have introduced and studied a toy model in $1d$ for anomalous transport and Griffiths effects in quantum disordered isolated systems near the Many-Body localization transitions. The model is build on random matrix theory and 
can be thought as a microscopic and analytically tractable realization of the coarse-gained effective models introduced in the framework of the strong 
disordered RG approach to MBL~\cite{vosk,potter1,thiery1,thiery,potter2}, 
and exhibits an intermediate Griffiths region where arbitrarily large ergodic and insulating segments coexist.
In particular, we have established the following key properties of the intermediate phase:
\begin{itemize}
\item The probability to find an insulating inclusion of size $L$ is exponential, Eq.~(\ref{eq:Ploc}). The presence of exponentially distributed localized regions is a distinctive feature 
of the Griffiths phase invoked to describe the properties of the bad metal phase close enough to the MBL transition.
\item The mechanism for delocalization is driven by proliferation of quantum avalanches~\cite{avalanches,thiery1,thiery,dimitrescu,goremykina,morningstar}, 
i.e., a network of few, rare resonances that destabilizes the insulating phase. 
This yields a broad distribution of the dissipation propagation of conducting samples, which can take any values in the interval $[e^{-L/\xi_{\rm typ}},1]$.
\item While typical correlations decay exponentially over a length $\xi_{\rm typ}$ which is proportional to the bare single-particle localization length $\xi_0$ of the $1d$ Anderson 
insulator and finite, average correlations decay as stretched exponentials, 
corresponding to a fractal structure of conducting inclusions. The fractal dimension $d_{\rm f}$ is found to decrease (slowly) with $L$;
\item This behavior is consistent with the KT-like criticality of the MBL transition~\cite{dimitrescu,goremykina,morningstar}, 
and can be interpreted in terms of a broadly distributed localization length, with a typical value $\xi_{\rm typ} \propto \xi_0$ and finite, 
and heavy power-law tails $\Xi(\xi) \sim \xi^{-2}$ at large $\xi$, such
that the average localization length is infinite at the critical point;
\item Transport and relaxation show anomalous behaviors strikingly similar to those observed in recent simulations~\cite{daveBAA,dave1,BarLev,demler,alet,torres,luitz_barlev,doggen,evers} and experiments~\cite{experiments1,experiments2,experiments3} in the bad metal delocalized phase preceding MBL. 
In particular, we find sub-diffusive transport and slow power-laws decay of the return probability at large times, with exponents that gradually change as one moves across the intermediate region. Concomitantly, the a.c. conductivity vanishes near zero frequency
with an anomalous power-law, and the distribution of resistivities of a fixed-sized sample grows increasingly broad at low-frequencies.
\end{itemize}
The analysis presented here 
yields a first step to bridge the gap between microscopic physics and long-wavelength critical behavior of the effective coarse-grained models developed in the 
context of the strong disorder RG approach for MBL~\cite{vosk,potter1,potter2,thiery,thiery1},
and provides a ``zeroth-order'' approximation for future studies.
A straightforward extension of the model consists for instance in adding some amount of disorder within each layer, by modifying the random potential of the first term
of~(\ref{eq:H}) as $\epsilon_i \to \epsilon_i + v_{i,p}$, with $v_{i,p}$ i.i.d in the interval $[-\Delta_i/2,\Delta_i/2]$ whose width $\Delta_i$ may vary along the chain.
In fact, if each layer $i$ is thought as a representation of a coarse grained block of length $\ell$ of an interacting model 
and the sites of the RRGs as 
many-body configurations of the coarse-grained degrees of freedom in some local basis, it is natural to assume that each local configuration should be associated to a random 
energy. The amplitude $\Delta_i$ of the intra-layer disorder should thus also be a random variable of with proportional to $W$. 
On the layers where $\Delta_i$ is large enough, the Anderson problem on the $i$-th RRG might be in the localized phase, and the intra-layer coupling might not be effective in
providing a local source of dissipation. 
Hence, depending on the specific realization of the random chemical potential on the corresponding interacting problem, 
some of the blocks $i$ might locally behave more 
like insulator or more like thermal systems.
This is very similar in spirit to the effective coarse-grained models developed in~\cite{vosk,potter1,potter2}, and possibly correspond to a more realistic description of real systems
close to the MBL transition.
 
As discussed above, characterizing the asymptotic KT-like critical behavior from the numerical analysis of finite size samples is a very hard task~\cite{morningstar}
In particular we find that $d_{\rm f}$ is affected by strong finite-size effects, and is still very far from the expected value $d_{\rm f} \to 0$ for the largest values of $L$ that we can access numerically.
We thus leave the precise determination of the critical exponents $\nu$ and $d_{\rm f}$ for future investigations~\cite{future}. Along the same line, it would be 
helpful to adapt and implement the approximate RG transformations of~\cite{vosk,potter2,potter1,thiery,thiery1} to our toy model to investigate the universality class of the RG-flow.
In Refs.~\cite{goremykina,morningstar} the relevant scaling variables of the KT RG flow have been identified as the density of thermal regions and the length scale that controls
the decay of typical matrix elements. In our toy model, we expect that the density of thermal inclusions is controlled by the ratio $L\gamma/W$, which gives the probability of finding a 
resonance in a chain of length $L$, Eq.~(\ref{eq:FSA}), while the internal length of insulating blocks that controls the decay of typical matrix elements should be proportional to $\xi_{\rm typ}$.
In this respect, it would be perhaps useful to exploit the formal equivalence of our model with the effective $1d$ Anderson Hamiltonian~(\ref{eq:Heff}) with complex and correlated random energies,
whose RG flow can possibly be worked-out exactly and up to very large sizes.

Several recent numerical and experimental works have pointed out that in fact quasi-periodic $1d$~\cite{experimentsQC1,experimentsQC2,qcs,mace,evers1} and 
disordered $2d$~\cite{2d} systems also display analogous unusual transport and relaxation, while on general grounds one expects that Griffiths effects should only give
a subdominant contribution when the potential is correlated and/or the dimension is larger than one~\cite{reviewdeloc1,griffiths2}. It is therefore natural
to seek for other mechanisms that might complement the Griffiths picture beyond the case of $1d$ disordered systems with uncorrelated disorder.
This was attempted in some recent works~\cite{PLMBL} where, using the Anderson model on the Bethe lattice as a
pictorial representation for the many-body quantum dynamics~\cite{dot}, an alternative explanation of the slow and power-law-like 
relaxation observed in the bad metal phase was proposed directly in terms on quantum dynamics in the Fock space.
Understanding the relationship between fractal Griffiths regions in real space and multifractality of the wave-function in the Hilbert space is certainly a very important problem.
Preliminary observations of the of energy levels and eigenvectors statistics of the present model in the intermediate phase seem in fact to suggest that 
a broad region of the phase diagram might be characterized by nonuniversal level statistics and multifractal wave-functions. We leave this analysis for future 
investigations~\cite{future}.

Finally, 
a parallel and promising line of investigation for future research is to analyze the properties of the metal/insulator transition of the model~(\ref{eq:H}) in the case of quasiperiodic potential 
either of the Aubry-Andr\'e or Fibonacci type.

\begin{acknowledgments}
We would like to thank D. Abanin, G. Biroli, L. Cugliandolo, I. V. Gornyi, F. Evers, L. Foini, D. Huse, G. Lemari\'e, A. D. Mirlin, M. M\"uller, G. Semerjian, M. Serbyn, K. S. Tikhonov, and V. Ros for many enlightening and helpful discussions.
Marco Tarzia is a member of the {\it Institut Universitaire de France}.
\end{acknowledgments}

\appendix

\section{Effective one-dimensional recursion} \label{app:eliminating}

In this appendix we discuss a possible route to eliminate the intra-layer couplings 
and to obtain an effective one dimensional model. Let's start from the last recursion for $G_i^{(v)}$, i.e.
$$
G_i^{(v)} = \frac{1}{-\epsilon_i-z-t^2G^{(l)}_{i+1}-t^2G^{(r)}_{i-1}-k\gamma^2G_i^{(v)}}
$$
if we define $w_i =-\epsilon_i-z-t^2G^{(l)}_{i+1}-t^2G^{(r)}_{i-1}$ we obtain a closed equation for $G_i^{(v)}$ which reads
\[
k\gamma^2 \left(G_i^{(v)}\right)^2- w_i \left(G_i^{(v)}\right)+1=0 \, ,
\]
from which we get the two branches
$$
\left(G_i^{(v)} \right)_{\pm} = \frac{w_i \pm \sqrt{w_i^2-4 k \gamma^2}}{2k\gamma^2} \, .
$$
We are going to choose the negative branch to make sure to recover the $\gamma\rightarrow0$ limit. Indeed if we plug this expression into the recursion for $G_i^{(l)}$, we obtain for example
\[
\begin{aligned}
\left(G_i^{(l)}\right)^{-1} = & -\epsilon_i-z-t^2G^{(l)}_{i+1} \\
& - \left(\frac{k+1}{2k}\right) \left(w_i-\sqrt{w_i^2- 4 k \gamma^2}\right) \, ,
\end{aligned}
\]
from which we recover the expected result for $\gamma=0$.  

We notice now that according to our definition $w_i$ is nothing but 
$$
w_i^{-1}=\frac{1}{-\epsilon_i-z-t^2G^{(l)}_{i+1}-t^2G^{(r)}_{i-1}}\equiv\mathcal{G}_i^{0}(G_i^{(l)},G_i^{(r)}) \, ,
$$
i.e. the diagonal resolvent in Eq.~(\ref{eq:recursion_final}) in absence of the coupling $\gamma$.  Therefore we can rewrite the exact recursion as
\begin{equation} \label{eq:recursion_sigma}
\left(G_i^{(l)}\right)^{-1} = -\epsilon_i-z-t^2G^{(l)}_{i+1}-\Sigma(G_i^{(l)},G_i^{(r)})
\end{equation}
with 
\begin{equation} \label{eq:sigma_corrections}
\Sigma= \frac{1}{\mathcal{G}_i^{0}} \left(\frac{k+1}{2k}\right)
\left(
1-\sqrt{1-4 k \gamma^2 \left(\mathcal{G}_i^{0}\right)^2}
\right) \, ,
\end{equation}
and a similar equation for $\left(G_i^{(r)}\right)^{-1}$.  We can now for example expand the square root in power of $\gamma^2$ and obtain, to the lowest order,
\[
\begin{aligned}
\left(G_i^{(l)}\right)^{-1} & = -\epsilon_i-z-t^2G^{(l)}_{i+1}-(k+1)\gamma^2\mathcal{G}_i^{0} \, , \\
\left(G_i^{(r)}\right)^{-1} & = -\epsilon_i-z-t^2G^{(r)}_{i-1}-(k+1)\gamma^2\mathcal{G}_i^{0} \, ,
\end{aligned}
\]
which are a set of closed equations for $G_i^{(l)}$ and $G_i^{(r)}$ given that 
\[
\mathcal{G}_i^{0}=\frac{1}{-\epsilon_i-z-t^2G^{(l)}_{i+1}-t^2G^{(r)}_{i-1}} \, .
\]
In other words, upon eliminating the cavity Green's function of the intra-layer degrees of freedom we have obtained an effective one dimensional recursion for the left/right cavity Green's functions. As opposed to the standard one-dimensional non-interacting Anderson problem, to which it reduces for $\gamma=0$, this recursion is highly non-linear, due to the presence of the \emph{self-energy} term $\Sigma(G_i^{(l)},G_i^{(r)})$, and couples together left and right cavity Green's functions. It is therefore tempting to interpret the net effect of the GOE perturbation in terms of an effective interaction for the longitudinal degrees of freedom.

\section{Perturbative expansion in $\gamma^2$ of the self-energy} \label{app:self}
The simplest way to do determine the transition point of the model~(\ref{eq:H}) is to determine the convergence of the perturbative expansion in $\gamma^2$ for the real part of the self-energy once the iteration relations~(\ref{eq:recursion}) have been 
linearized~\cite{abou}.
The (cavity) self-energies on a site $i$ are defined as:
\[
\Sigma_i^{(l,r,v)} = S_i^{(l,r,v)} + i \Delta_i^{(l,r,v)} \equiv -\epsilon_i - z  - \left[G_{i}^{(l,r,v)}\right]^{-1} \, .
\]
In the localized phase its imaginary part vanish for $\eta \to 0^+$.  Hence, close  to  the  localization
transition, one can take the limit $\eta \to 0^+$ from the start and linearize the recursive equations for the self-energy with respect to $\Delta_i^{(l,r,v)}$: 
\begin{equation} \label{eq:S_Delta}
\begin{aligned}
S_i^{(l,r)} &= - \frac{1}{\epsilon_{i\pm 1} + S_{i \pm 1}^{(l,r)}} - \frac{(k+1) \gamma^2}{\epsilon_{i} + S_{i}^{(v)}} \, , \\
\Delta_i^{(l,r)} & = \frac{\Delta_{i \pm 1}^{(l,r)}}{(\epsilon_{i \pm 1} + S_{i \pm }^{(l,r)})^2} + \frac{(k+1) \gamma^2 \Delta_i^{(v)}}{(\epsilon_{i} + S_{i}^{(v)})^2}  \, , \\
S_i^{(v)} &= - \frac{1}{\epsilon_{i+1} + S_{i+1}^{(l)}} - \frac{1}{\epsilon_{i-1} + S_{i-1}^{(r)}} - \frac{k \gamma^2}{\epsilon_{i} + S_{i}^{(v)}} \, , \\
\Delta_i^{(r)} & = \frac{\Delta_{i+1}^{(l)}}{(\epsilon_{i+1} + S_{i+1}^{(l)})^2} + \frac{\Delta_{i-1}^{(r)}}{(\epsilon_{i-1} + S_{i-1}^{(r)})^2} + \frac{k \gamma^2 \Delta_i^{(v)}}{(\epsilon_{i} + S_{i}^{(v)})^2} \, .
\end{aligned}
\end{equation}
The real part of the self-energies can be systematically expanded in powers of $\gamma^2$ as $S_i^{(l,r,v)} = S_{i,0}^{(l,r,v)}  + \gamma^2 S_{i,1}^{(l,r,v)} + \gamma^4 S_{i,2}^{(l,r,v)} + \ldots$:
\begin{equation} \label{eq:S_expansion}
\begin{aligned}
S_{i,0}^{(l,r)} &= - \frac{1}{\epsilon_{i\pm 1} + S_{i \pm 1,0}^{(l,r)}} \, , \\
S_{i,1}^{(l,r)} &= \frac{S_{i \pm 1,1}^{(l,r)}}{(\epsilon_{i \pm 1} + S_{i \pm 1,0}^{(l,r)})^2 } - \frac{(k+1)}{\epsilon_{i} + S_{i,0}^{(v)}} \, ,\\
S_{i,2}^{(l,r)} &= \frac{S_{i \pm 1,2}^{(l,r)}}{(\epsilon_{i \pm 1} + S_{i \pm 1,0}^{(l,r)})^2 } - \frac{S_{i \pm 1,1}^{(l,r)}}{(\epsilon_{i \pm 1} + S_{i \pm 1,0}^{(l,r)})^3} + \frac{(k+1)}{(\epsilon_{i} + S_{i,0}^{(v)})^2} \, \\
S_{i,0}^{(v)} &= - \frac{1}{\epsilon_{i + 1} + S_{i + 1,0}^{(l)}} - \frac{1}{\epsilon_{i - 1} + S_{i - 1,0}^{(r)}} \, , \\
S_{i,1}^{(v)} &= \frac{S_{i + 1,1}^{(l)}}{(\epsilon_{i + 1} + S_{i + 1,0}^{(l)})^2 } + \frac{S_{i - 1,1}^{(r)}}{(\epsilon_{i - 1} + S_{i - 1,0}^{(r)})^2 }  - \frac{k}{\epsilon_{i} + S_{i,0}^{(v)}} \, .
\end{aligned}
\end{equation}
These equations can be easily solved order by order.
In practice, we expanded $S_i^{(l,r,v)}$ up to the $6$-th order in $\gamma^2$ and injected them 
into the exact recursive equations~(\ref{eq:S_Delta}), to check whether the result obtained from the perturbative expansion is a solution of the exact equations up tp 
some small corrections.

Note that the first contribution to $S_{i,1}^{(l,r)}$ coincide with the first contribution to the imaginary part of the self-energies $\Delta_i^{(l,r)}$. One can actually show that order by order the corrections to the real part of the 
self-energies obey a very similar equation as the imaginary part, Eq.~(\ref{eq:S_expansion}). This observation indicates that the GOE perturbation plays essentially the role of a thermal bath.

\section{Further information on the properties of the intermediate region} \label{app:crit}

In this appendix we provide more details, plots, and information on the properties of the critical region ($\phi=1$), 
where arbitrarily large insulating and metallic regions coexist.

\subsection{Transmission amplitudes} \label{app:trans}

\begin{figure}
\includegraphics[width=0.48\textwidth]{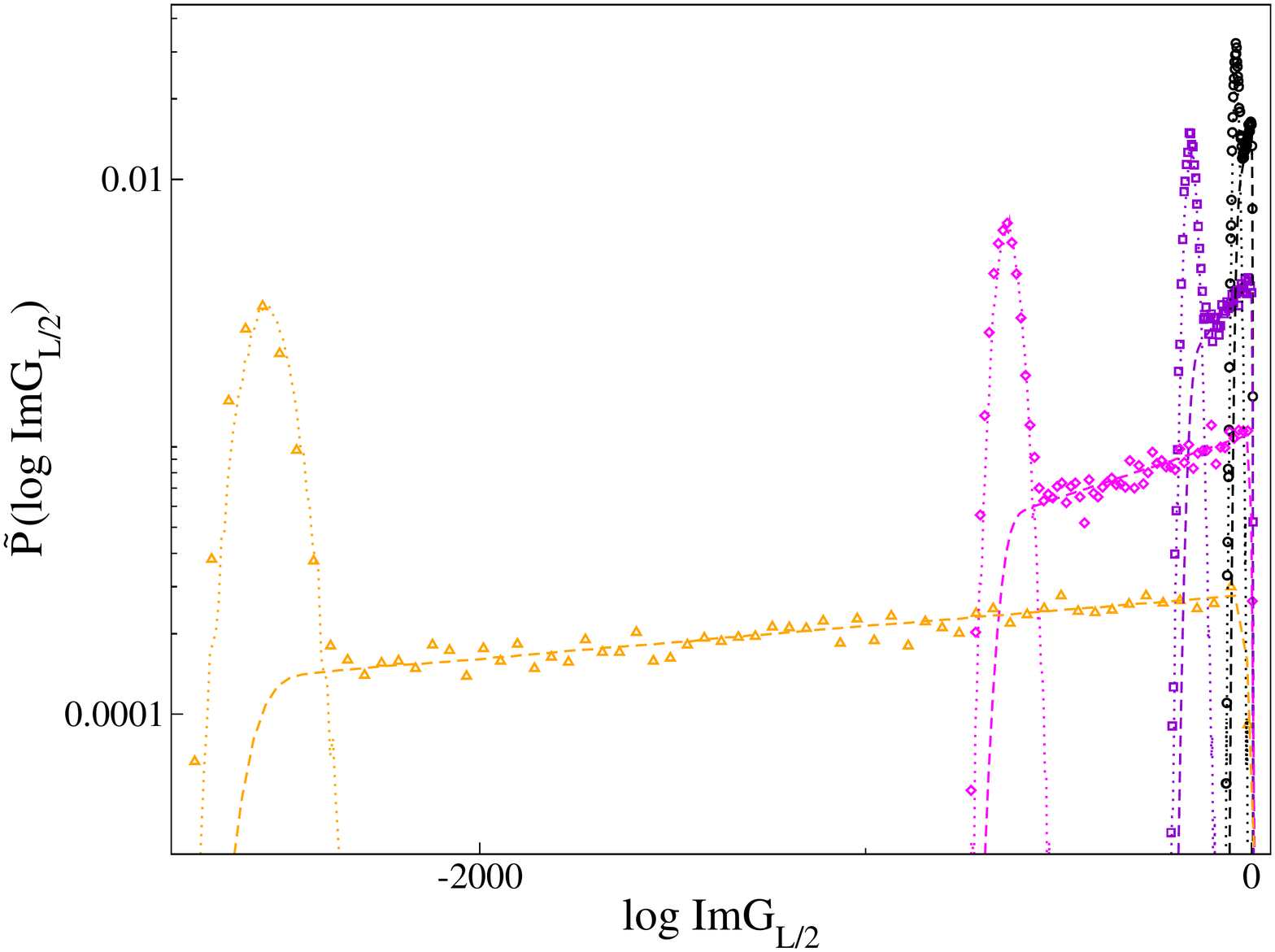}%
\caption{\label{fig:transmission}
Probability distribution (averaged over many realization of the disorder) of the log of the 
imaginary part of ${\cal G}_{L/2}$ in the middle of the system when $\eta$ is set identically to zero on all the layers of the chain except on the first layer, 
where $\eta=1$. The results are shown for $\phi=1$, $W=4$, $\hg=2$, and $L=2^8$ (black circles), $L=2^{10}$ (violet squares), $L=2^{12}$ (magenta diamonds), and $L=2^{14}$ (orange triangles). The dotted curves correspond to 
($P_{\rm loc}$ times) $\tilde{P}(\log {\rm Im} {\cal G}_{L/2})$ when the GOE-coupling is turned off ($\gamma=0$, i.e., for the standard Anderson tight-binding model in $1d$), while the dashed curves show ($1 - P_{\rm loc}$ times) the non-singular part of the
probability distributions $P(\log {\rm Im} {\cal G})$ plotted in Fig.~\ref{fig:plimg} when the
imaginary regulator is identically set to $\eta=0^+$ on all the sites of the chain.}
\end{figure}

We start by 
analyzing the probability distributions $\tilde{P}({\rm Im} {\cal G}_{L/2})$ of the imaginary part of the Green's functions at the middle of the chain, $x=L/2$, 
when the imaginary regulator is set to $\eta_i = \delta_{i,1}$, as in Sec.~\ref{sec:trans}.
Fig.~\ref{fig:transmission} shows $\tilde{P}({\rm Im} {\cal G}_{L/2})$
for $W=4$, $\hg=2$, and several values of the length chain $L$ (similar results are found for different values of $W$ and $\hg$ in the critical region).  
The probability distributions consists of two well distinct parts: 
The peak on the left, at very small values of ${\rm Im} {\cal G}_{L/2}$, corresponds to the insulating samples, for which the dissipation propagation 
decreases exponentially with the distance from site $1$ and no resonances are found. The peak is shifted to smaller and smaller values of ${\rm Im} {\cal G}_{L/2}$ as $e^{-L/(2 \xi_{\rm typ})}$
as the system 
size is increased and is essentially the same as the one that one would find in absence of the GOE coupling $\gamma$ (dotted curves), since $\xi_{\rm typ} \approx \xi_0$.
Conversely, the flat part of the distributions on the right corresponds to the conducting samples, for which the effect of turning on the intra-layer coupling is non-perturbative. This part of the distribution coincides
with the one shown in Fig.~\ref{fig:plimg}, when $\eta = 0^+$ on all the positions of the chain (dashed curves). It is essentially flat (since the value of ${\rm Im} {\cal G}_{L/2}$ for the conducting samples is set by the position of the closest resonance) 
and it stretches to lower and lower values of ${\rm Im} {\cal G}_{L/2}$ when the system size is increased [see Eq.~(\ref{eq:pimg})].
The area below the left part is $P_{\rm loc}$ and the area below the right part is $1 - P_{\rm loc}$, and changing the value of $\hg$ only changes the relative heights of the two parts.
This plots indicates that with probability $P_{\rm loc}$ the system is insulating and the transmission amplitude decreases exponentially fast as $e^{-L/\xi_{\rm typ}}$ (with $\xi_{\rm typ} \approx \xi_0$) as the system size is increased, while with probability $1 - P_{\rm loc}$ the system 
is conducting, yet the transmission amplitude is very small on most of the sites of the samples and of order $1$ only in the vicinity of few, rare resonances.
One thus expects that the distribution of the dc conductivity of a chain of length $L$ is also 
broad, 
with conductivities ranging from arbitrarily small values to values of order $1$. This is indeed confirmed by exact diagonalizations, as shown in App.~\ref{app:resist} and in Fig.~\ref{fig:psigma}.

\subsection{Statistics of the avalanche size distribution} \label{app:avalanches}

\begin{figure}
\includegraphics[width=0.48\textwidth]{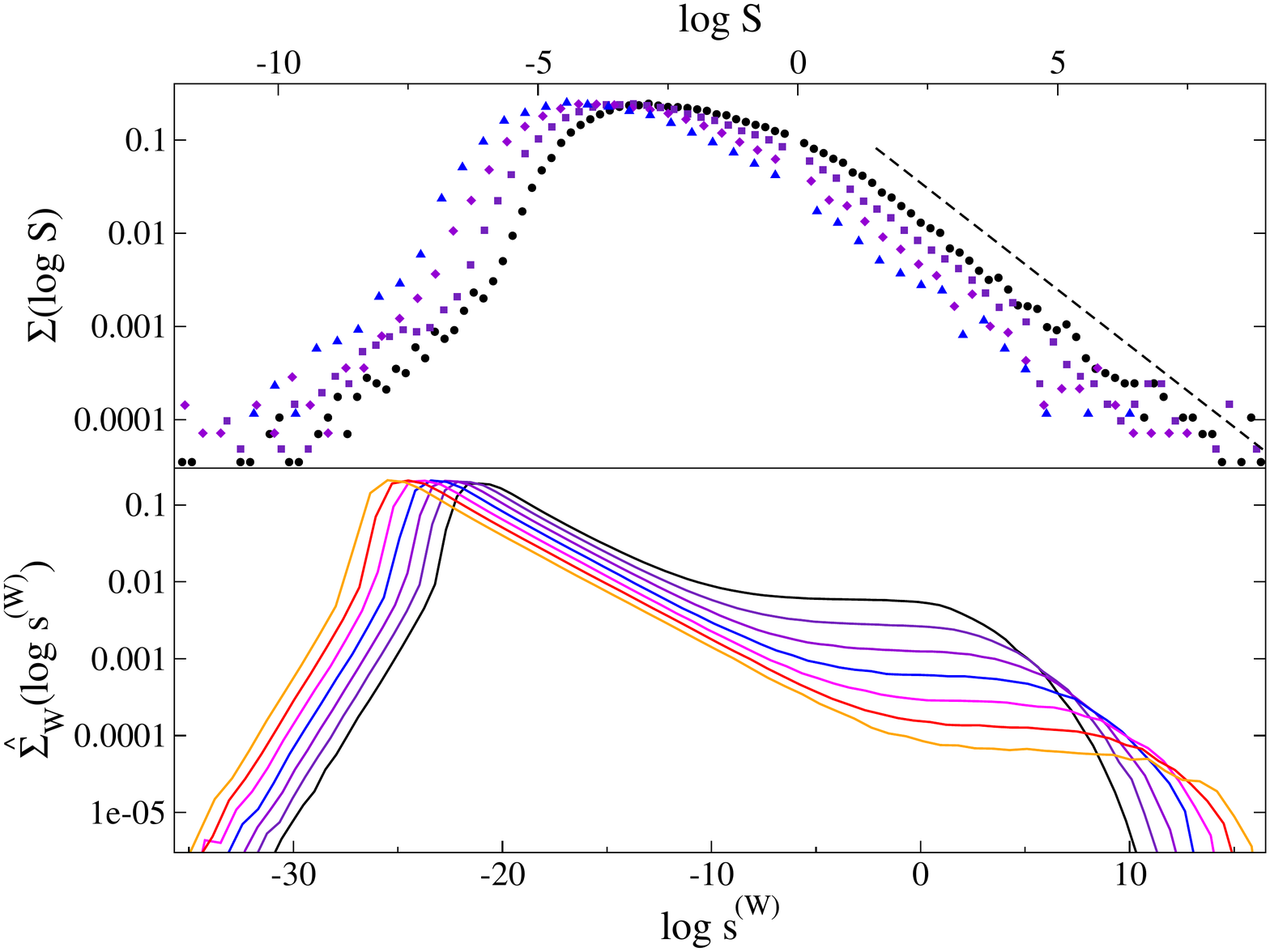}%
\caption{\label{fig:psusc}
Top panel: Probability distributions of the (logarithm of the) ``susceptibility'' 
$\Sigma( \log S)$ for $\phi=1$, $W=8$, $\hat{\gamma}=2$, and $L=2^8$ (black), $L=2^9$ (indigo), $L=2^{10}$ (violet), and $L=2^{11}$ (blue). The black dashed line
correspond to the power-law tails $\Sigma(S) \sim S^{-2}$. Bottom panel: Probability distributions of the (logarithm of the) ``local susceptibility'' 
$\hat{\Sigma}_W ( \log s^{(W)})$ for $W=8$, $\hat{\gamma}=2$, and $L=2^8$ (black), $L=2^9$ (indigo), $L=2^{10}$ (violet), $L=2^{11}$ (blue), $L=2^{12}$ (magenta), $L=2^{13}$ (red), 
and $L=2^{14}$ (orange).
The same behavior is found for $\hat{\Sigma}_{\hat{\gamma}} (s^{(\hat{\gamma})})$.}
\end{figure}

The statistics of the avalanche size distribution in the critical region observed in Fig.~\ref{fig:dega} when the coupling with the GOE-like perturbation is increased (see Sec.~\ref{sec:avalanches}) can be obtained by computing the ``susceptibility''
\[
S = \left . \frac{\partial \langle \log {\rm Im} {\cal G} \rangle}{\partial \hat{\gamma}} \right \vert_W\, ,
\]
which measures the increase of the (logarithm of the) typical value of the LDoS with respect to an infinitesimal increse of $\hat{\gamma}$ (at fixed $W$ and for a given realization of the disorder).
The probability distributions $\Sigma(\log S)$ are plotted in the top panel of Fig.~\ref{fig:psusc} for $\hat{\gamma} = 2$, $W=4$, and several systems sizes, showing that
$\Sigma(S)$ is broadly distributed, with heavy tails $\Sigma(S) \sim S^{-2}$ (black dashed line) which dominate the average: 
While for most of the samples
the ``response'' of the typical value of the  LDoS to an infinitesimal increase of $\hat{\gamma}$ is small (i.e., exponentially small in the length of the chain), there are few, rare samples for which a 
small increase of $\hat{\gamma}$ produces the formation of a new resonance and a sharp increase 
of the typical value of the  LDoS (i.e. a jump of order $L$ of $\langle \log {\rm Im} {\cal G} \rangle$).
The same behavior is found upon decreasing the disorder strength $W$ at fixed $\hat{\gamma}$.
Similarly, one can also define the ``local susceptibilies''
\[
\begin{aligned}
s_i^{(\hat{\gamma})} & = \left .\frac{\partial {\rm Im} {\cal G}_i}{\partial \hat{\gamma}} \right \vert_W \, , \\
s_i^{(W)} & = - \left . \frac{\partial {\rm Im} {\cal G}_i}{\partial W} \right \vert_{\hat{\gamma}} \, , 
\end{aligned}
\]
In Fig.~\ref{fig:psusc} we show the probability distributions $\hat{\Sigma}_W (\log s^{(W)})$ for $W=8$, $\hat{\gamma}=2$, and several system sizes.
The peak of the distribution is shifted to smaller and smaller values of $s^{(W)}$ when the system size is increased, while at large values of $s^{(W)}$ the distriubutions exhibit
an almost flat part which stretches to larger and larger values when $L$ is increased, which correspond to a power-law tail of the form $\hat{\Sigma} (s^{(W)}) \sim 1/s^{(W)}$:  
On most of the sites of the chain and for most of the samples 
the local DoS is left essentially unchanged by an infinitesimal decrease of the disorder strength as $\epsilon_i \to \epsilon_i(1 - \delta W)$, 
(i.e., the susceptibilities are exponentially small in $L$); Yet, for few specific samples which are at the brink of developing a new resonance, 
the response of the local DoS to an infinitesimal change of the disorder strength can be macroscopically large on the sites which are in the vicinity of the resonance.

\subsection{Lyapunov exponents} \label{app:lyap}

Further information on the anomalous critical behavior of the model can be obtained by studying
the probability distribution $L(\lambda)$ of the local Lyapunov exponents, which describe 
the exponential growth or the exponential decrease of the imaginary part of the Green's functions with the number of recursion steps $n$ when one starts from an infinitesimally small value: 
${\rm Im} {\cal G}_i^{(n)} \propto e^{n \lambda_i} {\rm Im} {\cal G}_i^{(0)}$. The most accurate way to compute the $\lambda_i$'s is provided by the ``inflationary'' algorithm put forward in Ref.~\cite{ioffe1}. The idea is to include 
an additional step to the recursion Eqs.~(\ref{eq:recursion}) where all the ${\rm Im} G_i^{(l,r,v)}$ are multiplied by a factor $e^{-\Lambda_n}$ so to keep the
typical imaginary part fixed and small: $ e^{\langle \log {\rm Im} G_i^{(l,r,v)} \rangle} = \zeta$. In practice one has to make sure that $\zeta$ is chosen in such a way that it is much smaller than the typical values of 
${\rm Im} G_i^{(l,r,v)}$ in absence of the inflationary step. 
As soon as a stationary distribution $P_\zeta ({\rm Re} G^{(l,r,v)}, {\rm Im} G^{(l,r,v)})$ is reached in this recursive procedure, the local Lyapunov exponents are defined from the exponential growth (or decrease) rate of 
${\rm Im} {\cal G}_i$ on the $i$-th layer of the chain between two iteration step: 
\[
\lambda_i \equiv \log {\rm Im} {\cal G}_i^{(n+1)} - \log {\rm Im} {\cal G}_i^{(n)} \, .
\]
This procedure also gives access to the global Lyapunov exponent $\Lambda$ associated to the exponential growth (or decrease) of the typical value of ${\rm Im} {\cal G}$ over the whole system: When the stationary distribution 
is reached on has that $\Lambda_n \to \Lambda$.

\begin{figure}
\includegraphics[width=0.48\textwidth]{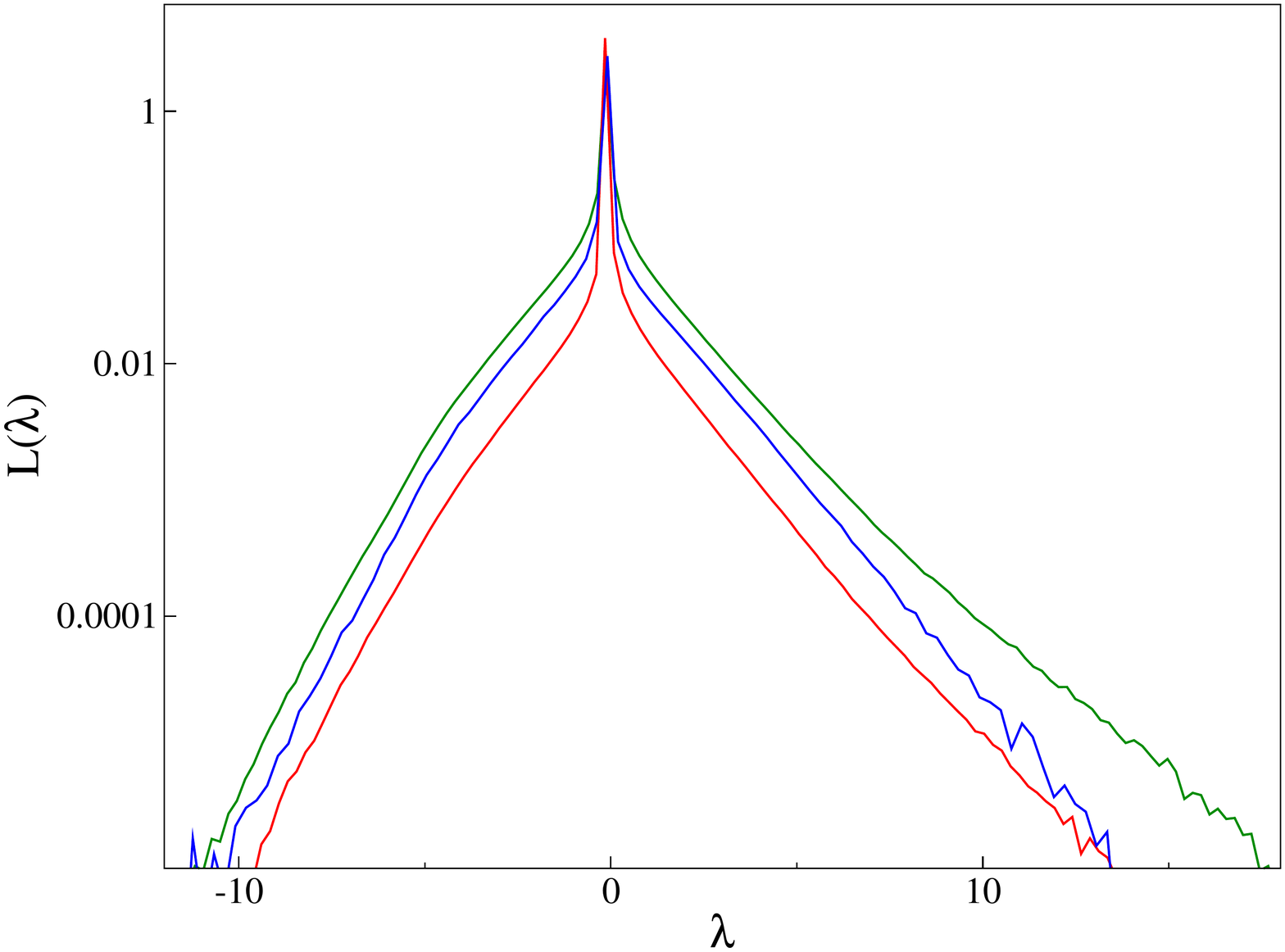}%
\caption{\label{fig:plyap}
Probability distribution of the local Lyapunov exponents $\lambda$ (averaged over several realizations of the disorder) 
for $\phi=1$, $W=1$, and $\hg=6$ (green), $W=4$ and $\hg=2$ (red), and $W=16$ and $\hg=2.7$ (blue), and for $L=2^{12}$.}
\end{figure}

The probability distributions of the local Lyapunov exponents for the samples which are in the conducting phase are shown in Fig.~\ref{fig:plyap} for three values of $W$ and $\hg$ and for $L=2^{12}$. The distribution functions show
a large peak corresponding to values of $\lambda_i$ close to $0$, and exponential tails on the left and on the right describing the behavior of $L(\lambda)$ at large and small $\lambda$ respectively: On the majority of the sites 
${\rm Im} {\cal G}_i$ grows slowly under iteration (when starting from infinitesimally small values), while there are few, exponentially rare, positions of the chains where ${\rm Im} {\cal G}_i$ grows or decrease very fast. 

Similarly, the probability distributions of the local Lyapunov exponents for the samples which are in the insulating phase (not plotted) exhibit a peak centered around a disorder-dependent value of $\lambda<0$ 
(which tends to zero for $W \to 0$), and exponential tails for negative values of $\lambda$. 

The presence of the exponential tails in the distributions $L(\lambda)$ is compatible with a quantum Griffiths phase~\cite{Vojta}, which is characterized by
exponential rare large insulating regions with exponentially large resistance.


\section{Power-law resistivity distribution in the Grffiths phase} \label{app:resist}
\begin{figure}
\includegraphics[width=0.48\textwidth]{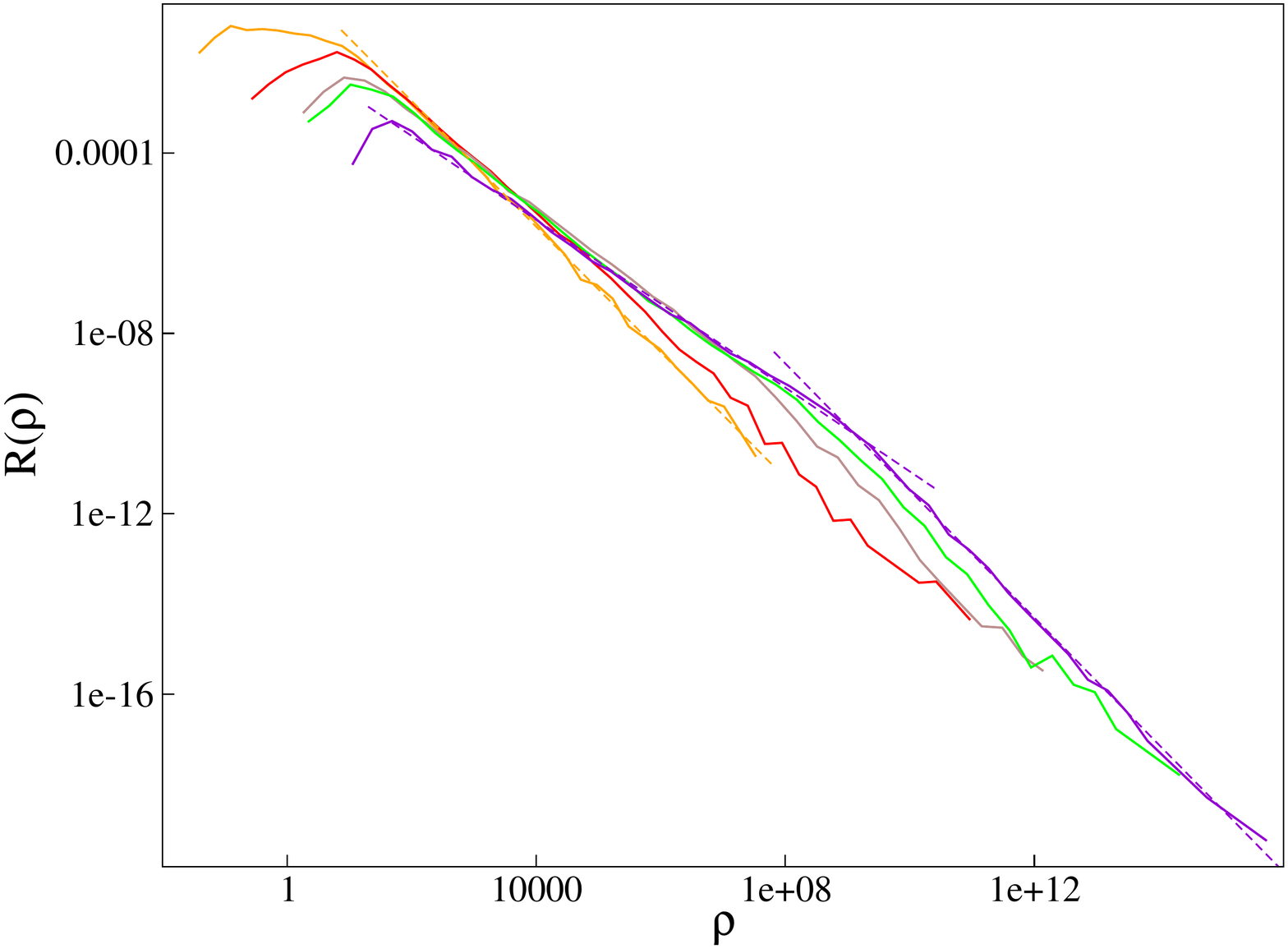}%
\caption{\label{fig:psigma}
Probability distribution of the resistivity $R(\rho)$ at a fixed sample size in the limit of low frequency $\omega \approx 1/(LM)$.
The data are obtained from exact diagonalizations of of finite-size samples with $L=44$ and $M=206$, and several values of the control parameter $\phi$ across the Griffiths region (light blue squares  in the interval $\phi \in [0.5,2.2]$ on the phase diagram of Fig.~\ref{fig:phase_diagram}). 
As $\phi$ is increased we observe a crossover between two power-law regimes, $R(\rho) \sim \rho^{- \tau^\prime}$ at intermediate values of $\rho$, 
with $\tau^\prime \approx 1$, followed by $R(\rho) \sim \rho^{- \tau}$ with $\tau \approx 1.5$ at larger values of $\rho$. The first power-law regime extends to
larger and larger values of $\rho$ as $\phi$ is increased, implying that also the average resistivity diverges in the $\gamma \to 0$ limit, recovering the properties
of the standard $1d$ Anderson insulator.}
\end{figure}

In this section we show the probability distribution of the resistivity $\rho=1/\sigma$ at a fixed sample size in the intermediate Griffiths phase, and in the limit of low frequency. 
The numerical results are plotted in Fig.~\ref{fig:psigma}, showing that the distribution approaches a power-law $R(\rho) \sim \rho^{-\tau}$. 
The authors of Ref.~\cite{demler} introduced a classical resistor-capacity network~\cite{RC} that 
reproduces some of the essential features of the Griffiths physics. On the basis of this model, it was suggested that $\tau = 2/(1+\alpha)$.
We indeed observe that $\tau$ slightly increases from $\tau \approx 1.3$ to $\tau \approx 1.6$ when increasing $\phi$, although we are unable to reliably 
extract the exponent $\tau$ directly from the data, owing to the difficulty of taking the dc limit in a finite system.

Such behavior implies  that the width (and sufficiently high moments) of the resistivity distribution diverges for $\omega \to 0$~\cite{demler}, and
is characteristic of a quantum Griffiths phase~\cite{Vojta}, in which
power-law correlations emerge due to the interplay between the exponential rareness of large insulating regions and their exponentially large resistence 
(see also Fig.~\ref{fig:plyap}).


\begin{thebibliography}{99}
\bibitem{localized_samples} The probability distributions $P({\rm Im} {\cal G})$ of the samples that are localized are singular in the $\eta \to 0^+$ limit and are essentially the same as the ones obtained for the standard $1d$ Anderson
tight-binding model for $\gamma=0$ (see also Fig.~\ref{fig:transmission}).
\bibitem{pgammaloc} In fact, assuming that $\hg_{\rm loc}$ is randomly distributed from one sample to another according to $p(\hg_{\rm loc})$, 
the probability that the system is localized is given by $P_{\rm loc} = \textrm{Prob}[\hg < \hg_{\rm loc}] = \int_{\hg}^\infty p(\hg_{\rm loc}) {\rm d} \hg_{\rm loc}$, implying that $p(\gamma_{\rm loc})$ is an exponential distribution with average $\hg_c$: $p(\gamma_{\rm loc}) = (1/\hg_c) e^{- \hg_{\rm loc} / \hg_c}$.
\bibitem{cusp} The cusps and the jumps are in fact rounded on the scale of the typical value of the 
LDoS (which is exponentially small in the system size), and become sharper and sharper as $L$ is increased.
\bibitem{largeM} Note that $L$ and $M$ should be chosen in such a way that the mean level spacing $\sim 4 \sqrt{k} \gamma/M$ of the local GOE-like perturbation stays 
much smaller than the mean level spacing $\sim W/L$ of the unperturbed single-particle Anderson localized levels. This implies that $q \gg 4 \sqrt{k} \gamma/ W$.
\end{thebibliography}
\end{document}